\documentclass[final,3p,times,twocolumn]{elsarticle}

\usepackage{graphicx}
\usepackage{amssymb}
\usepackage{amsthm}
\usepackage{amsmath}
\usepackage{mathrsfs}
\usepackage{lineno}
\usepackage{xcolor}

\def\etal{{\it et al.\/}}

\journal{Icarus}

\begin{document}

\begin{frontmatter}


\title{Large Mass Inflow Rates in Saturn's Rings due to 
Ballistic Transport and Mass Loading}



\author[IndU]{Richard H. Durisen}
\author[Ames,SETI]{Paul R. Estrada}

\address[IndU]{Department of Astronomy, Indiana University, Bloomington, IN 47405}
\address[Ames]{NASA Ames Research Center, Moffett Field, CA 94035}
\address[SETI]{Carl Sagan Center, SETI Institute, Mountain View CA 94043}

\begin{abstract}
 
The Cassini mission 
provided key measurements needed to determine the absolute age of Saturn's 
rings, 
including the extrinsic micrometeoroid flux at Saturn, the volume fraction 
of non-icy pollutants in the rings, and the total ring mass. 
These three factors 
constrain the ring age to be no more than {a few 100 Myr}
\citep{Kem23}. 
{Observations during} the Cassini Grand Finale also 
{showed that the} 
rings are losing mass to the planet at a prodigious rate. Some of the mass flux falls as 
``ring rain'' at high latitudes. 
However, {the influx in ring rain} 
is considerably less than the total measured mass influx of $4800$ to $45000$ kg s$^{-1}$ 
at lower latitudes \citep{Wai18}. 

In addition to polluting the rings, micrometeoroid 
impacts lead to {ballistic transport}, 
the mass and angular momentum transport due to net exchanges of 
meteoroid impact ejecta.
Because the ejecta are 
predominantly prograde, they carry net angular momentum outward. As a result, 
ring material drifts inward toward the planet. Here, for the first time, we 
{use a simple model to}
quantify this 
radial mass inflow rate for dense rings {and find} 
that, for plausible choices of parameters, ballistic transport and mass loading 
by meteoroids can produce a total inward flux of material 
in the inner B ring and in the C ring that is on the order of a 
few $\cdot\; 10^3$ to a few $\cdot\; 10^4$ kg s$^{-1}$, in agreement with
measurements during the Cassini Grand Finale. From these mass inflow rates, 
we estimate that the remaining ring lifetime 
is $\sim$ {15 to 400 Myr}. 
Combining this with a {revised} pollution age 
of $\sim$ {120 Myr}, 
{we conclude that} Saturn's rings are not only young but 
ephemeral and probably started their evolution 
{on a similar timescale to their pollution age}
with an initial 
mass of one to a few Mimas masses. 

{In addition to showing that meteoroid impacts can produce a large
sustained mass inflow through the B and C rings, this paper}
{addresses various} uncertainties and considers possible 
contributions by additional transport mechanisms and by external torques. 
We also map out a set of future research projects, {including global simulations.} 
 
\end{abstract}


\end{frontmatter}


\section{Introduction}
\label{sec:intro}

One of the impressive results of the Cassini Grand Finale was the measurement of a 
surprisingly large mass inflow into Saturn from the rings \citep{Wai18,Hsu18,Mit18}, 
as summarized in \citet{Per18}.
This inflow is at least an order of magnitude or more than the  
ring rain inferred from observations of  Saturn's upper atmosphere (see Section \ref{subsec:finale}). 
The Cassini mission overall, including the final orbits, has also improved our 
knowledge of parameters and processes critical to assessing how strongly 
meteoroid bombardment affects the rings. In particular, meteoroid bombardment causes 
the rings to evolve due to mass deposition by the meteoroids (mass loading, hereafter ML) and due to 
exchanges between different ring regions of mass and angular momentum through impact 
ejecta produced by the meteoroids (ballistic transport, hereafter BT). What we demonstrate in 
this paper is that these processes can plausibly produce a mass inflow through the B and C
rings toward Saturn of the magnitude observed.

This paper is organized as follows. Section \ref{sec:back} details the motivation for this work,
especially the large mass inflow into Saturn detected from the rings and the other new constraints 
on ring properties and environment derived from the Cassini Grand Finale. 
In Section \ref{sec:BT}, we review some basics of transport due to meteoroid bombardment.
In Section \ref{sec:inflow}, we develop simple models for the mass inflow expected from BT 
and ML, and we apply these models to Saturn's rings in Section 
\ref{sec:rings}, along with an estimate for the mass inflow rate due to viscosity.
Section \ref{sec:Discussion} presents a discussion of the mass influx rates, the absolute ring age,
 and the expected ring lifetime, and we summarize our conclusions and planned 
future work in Section \ref{sec:conclusion}. 

\section{Background and Motivation}
\label{sec:back}

\subsection{Ring Age and Pollution by Meteoroids}
\label{subsec:pollution}

Saturn's rings are continuously subjected to bombardment by extrinsic micrometeoroids. This affects the
rings in two principal ways. First, incoming impactors can lead to structural evolution of the rings 
because the copious ejecta produced by impacts cause mass and angular momentum exchanges between 
different ring regions 
{ \citep[][hereafter I83, I84, D84, D89, D92, D96, E15 and E18]{Ip83,Ip84,Dur84,Dur89,Dur92,Dur96,Est15,Est18}.} 
The meteoroids can also deposit mass and exert net
torques on a global scale {(I84; D96; E15; E18)}. Second, the predominantly icy rings become more and 
more polluted because the incoming material, which has a large non-icy component, darkens the rings 
over time \citep[][hereafter CE98]{CE98}. In particular, lower mass surface density $\Sigma$ 
and/or optical depth $\tau$ regions such as the C ring and Cassini division darken faster 
than the denser, more massive A and B rings. The susceptibility of the rings to the effects of 
micrometeoroid bombardment comes about in no small measure because of the rings' massive
surface-area-to-mass ratio, about a factor of $\sim 10^4-10^5$ times more than for a moon of 
equal mass (E15).

Key parameters, such as the rate of deposition of extrinsic material, the ring mass, and the current level of ring
pollution, as well as an understanding of the physics of how material is redistributed in the rings over 
time, provide a powerful age-dating tool for the rings. Prior to the Cassini 
Grand Finale, some of these parameters remained 
the subject of debate and speculation. That the rings might in fact be younger than previously thought was
first suggested by Voyager measurements showing that the albedo in the A and B rings were inconsistent 
with the fraction of pollutant they would have obtained over the age of the Solar System given the accepted 
micrometeoroid influx at the time \citep[][CE98]{Doy89}. The idea was considered unpopular because 
there was no apparent statistically plausible way to form the rings so long after the Late Heavy Bombardment 
if the rings were only a few hundred million years old \citep[see, e.g.,][]{Don91,Cha09}.
Thus, the view that the rings cannot be young persisted until Cassini could provide definitive evidence.

Analyzing Cassini radiometer measurements, \citet{Zha17a,Zha17b}
found that the non-icy material volume fractions in the C ring and Cassini division were $\sim 1-2\%$, whereas 
the fractions in the relatively more massive A and B rings were $\sim 0.1-0.5\%$ . These fractions are 
consistent with HST \citep{Cuz18} and VIMS results \citep{Cia19} if the pollutant is intramixed 
in fine grains and shadowing is allowed for. While visible and IR wavelengths only sample 
the surface layers of the particles, the microwave measurements sample the bulk of the ring particles, so 
that the radial variation of pollutants across the rings indeed indicates that the rings overall are not 
very polluted and, moreover, exhibit a distinct contrast in the level of pollution between low- and high-$\tau$ 
regions. Although low-$\tau$ regions darken faster than high-$\tau$ regions, this contrast diminishes 
over long times because the darker regions become the dominant source for darkening of the brighter 
higher-$\tau$ regions due to ballistic transport \citep[E15;][]{Est17}. While there is
evidence that the middle C ring is polluted with non-icy material to higher levels than the rest
of the C ring, \citet{Zha17b} showed that this can be explained by rubble from a captured 
Centaur that has become mixed into this region over the last few to ten million years or 
so. It does not affect the estimate of an overall pollution age.

In another caution about pollution ages, the Conclusions Section of \citet{Bur19} 
asserts that their evidence for icy E ring dust accumulating on the surfaces of outer ring moons
may invalidate young ring ages. Actually, what these authors find is that Pan 
in the Keeler gap has the same red color as the surrounding 
rings and that the reddening diminishes outward systematically 
as the distance of ring moons from Saturn increases. This suggests that the icy E ring dust,
or whatever else may be the source of the icy material, is being 
deposited only in the outermost ring region and does not penetrate as far as Pan. 
Pollution age analyses are not sensitive to the outer few
thousand kilometres of the rings but rely on comparisons between optically thick and thin
regions much closer to Saturn. The orbital dynamics of E ring dust 
\citep{Ham93err,Ham93,HB94} make it unlikely that it would 
penetrate deeply into the main rings. So, while a refined treatment of pollution
should probably consider sources of icy material, it is unlikely to affect ages
based on meteoroid pollution for the bulk of the rings
(see, however, Section \ref{subsubsec:compinflow}).
  
Another key piece of data for estimating pollution ages is the
input rate from meteoroids.
Several authors estimated the mass flux of meteoroids at Saturn prior
to Cassini \citep[e.g.,][hereafter CD90]{Mor83,Gru85,CD90}. After a careful
review of this work, CE98 showed in their Figure 17 that their improved 
estimate (the solid histogram in the figure) agreed with Pioneer and Ulysses 
data for 10 $\mu$m sized particles and smaller. They concluded that the CD90 
value for a one-sided interplanetary flux of 4.5 $\cdot$ 10$^{-17}$ g cm$^{-2}$ s$^{-1}$ 
for small meteoroids at Saturn, the value used in ballistic transport 
and pollution calculations up to that time, was good
to within a factor of three up or down. 
{Here, ``one-sided" means the flux measured passing through 
an area only from one side to the other.}
The Cassini Cosmic Dust Analyzer (CDA) 
experiment has now measured the meteoroid flux at Saturn and infers
an interplanetary flux at 10 AU {a factor of two less than the CD90 value} 
\citep[][see Section \ref{subsec:parameters}]{Alt15,Alt18,Kem17,Kem23}.

Furthermore, 
CDA finds, through orbit reconstruction, that the population 
of impactors is consistent with Edgeworth-Kuiper Belt (EKB) projectiles, not the Oort Cloud source that had 
been previously assumed (CD90; CE98). Particles from this dynamical population have much lower 
velocities as they enter Saturn's Hill sphere (which we call ``at infinity'') than cometary particles, 
and thus are about ten times more focused gravitationally by the planet. In other words,
for a given micrometeoroid flux at infinity, the impact flux on the rings is an order of magnitude 
higher than previously assumed, thus lowering age estimates by about the same factor. 
This is illustrated by Figure 9.5 of E18, where the computation of CD90 was redone using 
a meteoroid velocity at infinity at Saturn of 3 km s$^{-1}$ instead of the 14.4 km s$^{-1}$
used in CD90. 
The increase in the ejecta intensity is an order of magnitude\footnote{The actual mean velocity at infinity as derived by CDA {is} 
$4.3$ km s$^{-1}$ lowering the estimate of \citet[][who estimated a factor of $\sim 25$ higher]{Est18} 
by a factor of $\sim 2-3$, but still overall an order of magnitude in ejecta intensity.}. A similar factor 
can be obtained by applying the various gravitational focusing formulae
referenced in Appendix B of CE98. All these computations involve approximations. A more careful
treatment of gravitational focusing is needed but would be a substantial project. 
However, the order of magnitude difference in focusing between
the two meteoroid populations is almost certainly valid and has been accepted
for some time \citep[e.g.,][]{Mor83}.

Finally, the total ring mass plays a critical role in determining the exposure age of the rings. The 
post-Voyager estimate for the rings was roughly a Mimas mass \citep{Esp84}, but if the ring mass 
surface density, say in the opaque, dense B ring core were an order of magnitude higher -- an idea that 
seemed plausible from discovery of the rings' clump-and-gap structure in which density waves could hide 
large amounts of mass \citep{Ste07,Rob10} -- then the rings would be much more 
resistant to pollution and could be ancient \citep{Esp08}. However, from gravity field measurements 
in the final orbits of the Cassini Grand Finale, the mass of the rings has been determined to be 
$\sim 0.4$ Mimas masses \citep{Ies19}, even lower than the post-Voyager estimate and similar to 
estimates from wavelet-based analyses of several density waves in the B ring \citep{HN16}. 

Taken together, the direct bulk pollution measurements, the characterization of the meteoroid impactors,
and the determination of the ring mass unambiguously suggest a youthful age for the rings. 
As discussed in detail by E18, with the new observational constraints, 
the ages deduced for the rings from meteoroid bombardment based on various global, structural, 
and pollution considerations all seem to agree on an age on the order of at most a few $\cdot\; 10^8$ yrs. Rings  
with ages $\lesssim$ few $\cdot\; 10^8$ yrs {{are consistent with}} 
{{the ring age calculated by \citet{Kem23}.}}
 
\subsection{Ring Rain, Mass Inflow, and the Grand Finale}
\label{subsec:finale}

The end of the Cassini mission provided a unique opportunity to determine additional constraints for 
the rings. Observations conducted in its final 22 orbits, when the spacecraft flew through
the 2000 km-wide region that separates the rings from Saturn's upper atmosphere, measured the amount of  
ring material falling into the planet, thereby providing another means by which to constrain not only 
how old the rings are but also estimate how long their remaining lifetime may be.
These final orbits also allowed for the determination of the composition 
of the ring material and to test the so-called ``ring rain'' hypothesis. In this scenario, some fraction 
of the ejected material resulting from micrometeoroid impacts (in the form of ions or 
nanometer-to-sub-micron-sized grains), which can be charged either by micrometeoroid impact-produced 
plasma or photoionization, travel along magnetic field lines and fall into Saturn's (preferentially 
southern) atmosphere \citep{Con86,NC87,JH12,Ip16}. 
The transport of this ejected material, presumed to be mostly 
water, into the Saturnian atmosphere is thought to explain the latitudinal pattern of ionospheric 
H$_3^+$ infrared emission \citep[e.g.,][]{Odo13,Moo15}. 

Cassini's Magnetospheric Imaging Instrument (MIMI) which measured neutral atoms, ions and electrons was 
deployed, in part, to test the ring rain hypothesis. MIMI detected a very small flux of $\sim 1-2$ nm radius
grains (assuming the density of ice) precipitating into Saturn's atmosphere through the process of 
atmospheric drag due to kinetic collisions with exospheric H and H$_2$. Modeling of the data indicate that 
the grains, likely originating from the D68 ringlet at the inner edge of the D ring and  
mostly confined within $\pm 2^o$ of the ring plane, precipitate at a rate of $\sim 5.5$ kg s$^{-1}$ 
\citep{Mit18}. These authors also find that the charged component of these grains is relatively 
low in the midplane, but reaches a maximum in the wings of the distribution at higher latitudes 
($\pm 5^o$) consistent with transport along magnetic field lines associated with ring rain. This flux 
is small, but may not be surprising given that MIMI only samples the low end of the particle size 
distribution, whereas larger particles dominate the mass density \citep{Hsu18}. 

The CDA, which is sensitive to larger particles\footnote{{ The CDA instrument has multiple impact targets that allow for sensitivity to a range of particle size \citep[see][]{Sra04}.}}, was also active in the final orbits and
found that the primary flux into the planet arises from grains $\gtrsim 10-40$ nm in radius entering at
moderate latitudes. Larger grains (hundreds of nm) are also seen confined to a band within a
few hundred km of the midplane and with a lower abundance \citep{Hsu18}. Their simulations indicate 
that the observed population of nanograins are fast ejecta released from the surfaces of the rings due to
micrometeoroid impacts. These nanoparticles travel along magnetic field lines and are deposited in the 
atmosphere over a latitudinal range of $\pm 40^o$. \citet{Hsu18} estimate an ejection rate for these tens 
of nanometer-sized particles to be 1800 to 6800 kg s$^{-1}$ of which 18\% arrive at Saturn as ring rain, 
or roughly 320 to 1200 kg s$^{-1}$. Of this fraction, 70\% is deposited near the midplane, while 30\% 
(100 to 380 kg s$^{-1}$) is deposited in the mid-latitude ring rain regions.

The measured mass inflow rate is over an order of magnitude more than the 3 to 20 kg s$^{-1}$ of water product
initially thought to be needed to account for the H$_3^+$ emission \citep{Moo15}. However, a new 
analysis of the ground-based Keck II data \citep{Odo13}, in which the temperature, density, and
cooling rate parameters for H$_3^+$ were directly derived, now estimates that the required mass flux is 
432 to 2800 kg s$^{-1}$ of charged water products \citep{Odo19}. This is interesting in that the composition 
of the nanograins measured by CDA were found to be that of water and silicates, with the measured silicate 
mass fractions ranging from 8 to 30\% and the silicate fraction being maximum approaching the ring rain 
latitudes. Since the rings are by mass $> 95$\% water ice, the depletion of water in the CDA data suggests a mechanism
at work to erode water ice from the grains, but the ultimate fate of that water is not clear. These recent
results suggest that the water lost from these grains may indeed be ending up in the atmosphere\footnote{The
measured mass fraction in the midplane is close to C ring composition in terms of volume fraction ($\sim 3$\%) as
derived by \citet{Zha17b}.}.
 
The Cassini Ion Neutral Mass Spectrometer (INMS) experiment, which measured the composition of 
Saturn's equatorial upper atmosphere and its interactions with material originating from the rings, found 
a surprisingly large mass influx of between 4800 and 45000 kg s$^{-1}$ within a latitude band 8$^o$ 
near the equator \citep{Wai18}. In addition to observing the infall of water, substantial amounts of 
volatiles and impact fragments of organic nanoparticles were also measured. These authors concluded that 
this large mass flux likely requires the regular transfer of material from the C ring to the D ring.
Unlike CDA and MIMI, which were sensitive to a specific range of particle radii, INMS sampled
remnants of infalling material regardless of original particle size and so may 
offer a more appropriate measure of the total mass influx 
from the rings into the planet. The minimum and maximum values of the equatorial mass flux, which were 
derived from three low altitude revs, are significantly larger than the influxes estimated by MIMI and
CDA and larger than the amount of influx needed to explain the ring rain effect. 

It is the source of this large overall mass influx that we endeavor to explain in this paper
as a natural by-product of outward angular momentum transport  by BT and of meteoroid
mass loading. Specifically, our aim is, for the first time, to quantify the inward drift of ring 
material from the main rings into regions close to the planet due to micrometeoroid
bombardment and to determine how this compares with the overall
mass influx of material measured in the final orbits of the Cassini Grand Finale. If they
are comparable, it would provide another confirmation of the importance of meteoroid bombardment
for ring dynamics and a consistency check for age dating the rings.  
 
 \section{Ballistic Transport and Mass Loading}
\label{sec:BT}

Our own theoretical treatment of ballistic transport and mass loading
in Saturn's rings was developed 
in a series of papers over two decades ago (D84; D89; D92; D96) and was recently reviewed 
in depth by E15 and E18. The reader should consult these references for details beyond
the summary provided in this section.

Our formulation of the response of a ring to meteoroid bombardment, as
described below, only applies to dense rings, like the A, B, and C rings
of Saturn, and not to sparse or gossamer rings like the D ring, where other processes
are likely to dominate the behavior \citep{Hed18b}. We discuss the D ring and the problem of 
spanning the gap between the D ring and Saturn in Section \ref{subsubsec:CDrings}. The 
purpose of this paper is to demonstrate that ballistic transport and mass loading can sustain 
mass inflow toward Saturn through the B and C rings of the magnitude observed flowing onto Saturn.

\subsection{Basics}
\label{subsec:basics}

Saturn's rings are subject to impacts by small meteoroids, with
the bulk of the incoming mass in the form of micron-to-tens-of-micron-sized particles 
of interplanetary origin \citep[CD90;][]{Kem17,Alt18,Kem23}. The dust strikes the ring particles at 
hypervelocity with speeds of 
10's km s$^{-1}$. The primary result of impact is release of a high yield 
of small ejecta particles with speeds relative to their ring particle of origin 
that range up to 100's m s$^{-1}$ or more. This is true whether the impacts are
cratering events or cause disruption of the ring particle. The fragments of 
a disrupted ring particle can formally be treated in BT as ejecta. The ejecta 
that escape their radius of origin arc above the rings on ballistic elliptical 
trajectories which carry them to ring regions separated in radius 
by a few to a few thousands of km. Depending on the 
relative optical depths of the regions where the ballistic trajectories intersect the
ring plane, some or all of the ejecta are absorbed by the rings at a radius
different from their radius of origin. 

A characteristic time scale associated with BT, called the ``gross erosion time" 
(D84), is defined as the time it would take a ring region
to erode away completely if all locally produced ejecta were lost and no ejecta 
were gained from elsewhere. It is given by

\begin{equation}
\label{equ:teegee}
t_{\rm{G}} \equiv \frac{\Sigma}{\dot{\sigma}_{\rm{ej}}} \approx \frac{\Sigma}{<Y>\dot{\sigma}_{\rm{im}}},
\end{equation}

\noindent
where $\Sigma$ is the surface mass density of the rings at radius $r$ 
from the center of Saturn, $\dot{\sigma}_{\rm{ej}}$ is the (two-sided) mass emission rate of
ejecta per unit area, $\dot{\sigma}_{\rm{im}}$ is the (two-sided) mass rate of meteoroid
impacts per unit area, and $<Y>$ is {the typical yield, 
i.e., the typical ratio} of ejected mass to meteoroid mass for an impact. ``Two-sided" here means 
that inflow onto both sides of the rings, North and South, is included. As 
discussed below, $t_{\rm{G}}$ is on the general order of 10$^4$ to 10$^5$ yr. 
This is much longer than the local collision time for ring particles, typically a half to 
several orbit periods (hours to days), but short compared to the
Solar System lifetime. 

BT usually manifests itself on a time scale longer than $t_{\rm{G}}$ because 
not all ejecta from $r$ are lost and ejecta are also absorbed from other ring 
regions. It is the net effect of ejecta exchanges between ring regions 
that produces transport of mass and angular momentum. Net mass exchange 
leads to what we call a ``direct" change in the surface mass density 
of the rings. In addition, because the ejecta generally have specific angular momentum 
different from their parent ring particles and also different from the ring particles that absorb
them, ejecta exchanges cause the specific orbital angular momentum of the local 
ensemble of ring particles to change. The result is a slow drift to a different ring 
radius. The divergence of this local mass flux also changes the ring surface density.

The following equation from E18 describes the result of these two effects on the 
evolution of the ring surface density $\Sigma$, 

\begin{equation}
\label{equ:mass}
\frac{\partial{\Sigma}}{\partial{t}} = - \frac{1}{r}\frac{\partial}{\partial{r}}(\Sigma 
rv_{\rm{r}}) + \Gamma_{\rm{m}} - \Lambda_{\rm{m}} + \dot{\sigma}_{\rm{im}}.
\end{equation}
 
\noindent
Here we have adopted the same notation found in E18. 
$\Gamma_{\rm{m}}$ is the gain rate of mass per unit area 
at $r$ due to absorption of ejecta from elsewhere, and $\Lambda_{\rm{m}}$ is the
loss rate of mass per unit area due to ejecta emitted at $r$ and absorbed 
elsewhere. The first term on the RHS of Eq. (\ref{equ:mass}) is the change 
in $\Sigma$ due the divergence of the radial mass flux. 
$\Gamma_{\rm{m}} - \Lambda_{\rm{m}}$ represents the net local direct gain
of mass per unit area due to ejecta exchanges. The last term on the RHS
of (\ref{equ:mass}) is the mass loading, the mass deposited locally by 
the meteoroids themselves. Here we assume no meteoroid mass is lost.

The radial drift speed $v_{\rm{r}}$ in Eq. (\ref{equ:mass}) 

\begin{equation}
\label{equ:vr}
v_{\rm{r}} = v^{\rm{ball}}_{\rm{r}}+v^{\rm{visc}}_{\rm{r}}
+v^{\rm{load}}_{\rm{r}}
\end{equation}

\noindent
is caused by at least three processes, namely, from left to right, ballistic transport, 
viscosity, and mass loading. We will not concern ourselves here with the viscous
radial drift $v^{\rm{visc}}_{\rm{r}}$. An estimate for viscously driven mass inflow
will be given in Section \ref{subsec:visc}. 
The radial drift due solely to ejecta exchanges is given by 

\begin{equation}
\label{equ:vrball}
v^{\rm{ball}}_{\rm{r}} = \left[\Sigma \frac{dh_{\rm{c}}}{dr}\right]^{-1} \left
[\Gamma_{\rm{h}}-\Lambda_{\rm{h}}-h_{\rm{c}}(\Gamma_{\rm{m}}-\Lambda_{\rm{m}})
\right],
\end{equation}

\noindent
where $\Gamma_{\rm{h}}$ and $\Lambda_{\rm{h}}$ are the gain and loss rates of
angular momentum per unit area at $r$ due to ejecta gained and lost and $h_{\rm{c}}$ is the
specific orbital angular momentum at $r$ 
required for a circular orbit around Saturn. Basically, it
is assumed that, as its specific orbital angular momentum changes, the ensemble of 
ring particles at $r$ drifts  to a radius where its new value of $h_{\rm{c}}$
allows it collectively to follow a circular orbit.
Note that $\Gamma_{\rm m}$ and
$\Lambda_{\rm m}$ have different units from $\Gamma_{\rm h}$ and $\Lambda_{\rm h}$,
g cm$^{-2}$ s$^{-1}$ and g s$^{-2}$, respectively.

The radial drift due to mass loading and associated torques is given by

\begin{equation}
\label{equ:vrload}
v^{\rm{load}}_{\rm{r}} = \left[\Sigma \frac{dh_{\rm{c}}}{dr}\right]^{-1} 
\left[\dot{J}_{\rm{im}} - h_c\dot{\sigma}_{\rm{im}}\right],
\end{equation}

\noindent
where $\dot{J}_{\rm{im}}$ is the torque caused by the meteoroids and is 
usually negative due to aberration of the meteoroids by the orbital motions of Saturn and of the ring 
particles (CD90). Note that, even if there were no aberration, {\it i.e.}, $\dot{J}_{\rm{im}} = 0$, 
the meteoroid mass loading by itself will cause inward radial drift. In fact,
mass loading tends to dominate the meteoroid torque at least for Oort cloud projectiles (see D96).

\subsection{The Ejecta Distribution}
\label{subsec:ejecta}

There are a few features of the expected angular and speed distributions of 
ejecta that are worth noting before we consider approximations to the
gain and loss integrals. 

{\it Predominance of Prograde Ejecta.} 
Because of the aberration of the meteoroid
influx by the ring particle orbital motion (CD90), impacts tend to occur 
on the leading hemispheres of ring particles. So
the ejecta velocities from non-disruptive impacts tend to have a prograde
sense with respect to the ring particle orbital velocity. The vertical 
thickness of the ensemble of ring particles is only tens of meters \citep{Col09} 
with random speeds of less than a cm s$^{-1}$, much smaller than the ejecta
speeds. So, as discussed in D84, D89, and D92, prograde ejecta follow 
elliptical orbits that arc above the ring plane and re-intersect it at a 
larger radial distance $r_{\rm{int}}$. Prograde ejecta carry away more specific 
angular momentum than required for a circular orbit at $r$ but arrive 
at $r_{\rm{int}}$ with less specific angular momentum than needed there for a 
circular orbit. Retrograde ejecta have the opposite signs of angular 
momentum differences. Because the cratering ejecta are predominantly 
prograde, they represent a net outward transport of angular momentum. However,
both prograde ejecta lost from $r$ to $r_{\rm{int}}$ and those gained at $r$ 
from an interior radius decrease 
the specific angular momentum at $r$ and $r_{\rm{int}}$, and so cause an 
inward drift at both locations.
 
As shown in Fig. 12 and Eqs. (41) to (46) of CD90, 
the prograde ejecta outweigh retrograde ejecta by a considerable
factor when the impactors are predominantly Oort Cloud cometary dust.
The ratio can be seen to be about 2.5 to 1 for the approximations 
used in Fig. 3 of \citet{Lat12}.  The Cassini Dust Analyzer (CDA) 
results \citep{Kem17,Alt18,Kem23} show that the bulk of impactors on Saturn's rings 
are not cometary in origin but have the dynamics of
dust originating in the Edgeworth-Kuiper Belt, {\it i.e.}, low-to-moderate 
eccentricity and low, prograde inclinations in the heliocentric frame. 
Unlike the cometary population that is
isotropic in the heliocentric frame, the EKB population tends to be isotropic in the frame
of the planet and is characterized by a much lower speed $v_\infty$ at the Hill radius of Saturn
compared to Oort Cloud projectiles. As a result,  the EKB micrometeoroids 
are much more gravitationally focused
at the rings (Section \ref{subsec:pollution}). E18 redid the CD90
analysis for EKB impactors in an approximate manner 
and found that the resultant angular
distribution of the ejecta is even more prograde than the cometary one, with a 
ratio of prograde to retrograde ejecta of more like 7 to 1 (see Figs. 9.4
and 9.5 of E18), thus accentuating the radial inward drifts.

{\it Power-Law Speed Distribution.}
We characterize ejecta speeds $v_{\rm {ej}}$ by the parameter 
$x = v_{\rm {ej}}/v_{\rm {c}}$, where $v_{\rm {c}}$ is the local circular orbit speed.  
As discussed in E15 and E18, the speed distribution from non-disruptive
hypervelocity impacts is likely to be a steep power law $x^{-n}$ 
with $n$ between 2 and 3, and the speeds in this distribution
are likely to span the range from a few to up to several 100 m s$^{-1}$. 
For example, the cumulative ejecta speed distributions 
from laboratory experiments \citep{HH11} suggest values of 
$n \sim$ 2.1 to 2.7 for a wide range of target strengths 
(higher $n$'s) and porosity (lower $n$'s), 
and the hypervelocity impact experiments of \citet{KG01a,KG01b} using 
ice and ice/silicate targets produce maximum velocities of order several 100 m s$^{-1}$.

As shown by the $n =$ 2 and $n =$ 3 curves in Figure 9.10 of E18,
such $n$-values combined with a speed distribution ranging up to a few$\cdot$100 m s$^{-1}$ are 
probably required to produce the observed features called ``ramps",
which are linear in normal optical depth $\tau$ and exist just inside the steep 
A and B ring inner edges
(see Figs. 9.1, 9.2, and 9.10 of E18). Production of ramps for $n$ = 3 was also
discussed extensively in D92 and E15. 

One 
possible complication is that \citet{KG01a,KG01b} found 
the maximum velocity of ejection depended on the angle of ejection. 
CD90 included an ejecta cone model for impacts, including how the cone varied with 
angle of incidence of impact. The CD90 results ended up being averaged over
all possible impacts, and much of this detail washes out. 
CD90 did not include variation of the ejecta speed with
ejection angle. This could be done but would wash out by the time
impacts at all incidence angles were considered. Thought should be 
given to effects like this when the CD90 computations are redone in the
future, but what probably matters more is the average overall ejecta velocity 
distribution. The BT mass inflow results of 
Section \ref{subsec:inflowrates} are more sensitive to the lower ejecta 
speed cutoff of the power law, not the upper cutoff.
The upper cutoff matters more for explaining the extent of the
observed ramps.

Although disruptive 
impacts are also likely to occur, we have not yet tried to
characterize them or their BT implications in detail. Such impacts will probably 
include a cratering component of high-speed ejecta from the initial impact plus a 
high yield of slower moving, possibly more retrograde ejecta due to the breakup of
the ring particle. In this paper, we focus on the higher-speed prograde
yield from cratering events, which in fact will be most effective at
producing radial mass influx in the rings. We offer the ramps as 
observational evidence that a prograde power-law component 
does dominate BT at higher speeds.  
 
\subsection{The Gain and Loss Integrals}
\label{subsec:gainloss}
 
The gain and loss terms in Eqs. (\ref{equ:mass}) and (\ref{equ:vrball}) are given
by Eqs. (13), (14), (16), and (17) of E15. We describe them in this subsection only qualitatively. 
They involve triple integrals, two integrals over the solid angle of the angular ejecta 
emission function and one over the ejecta speed distribution. The emission angles 
and speeds determine, for the $\Gamma$'s, where ejecta that are absorbed at $r$ have 
come from and, for the $\Lambda$'s, where ejecta emitted at $r$ re-intersect the
ring plane. In addition, these integrals include a probability function that uses the optical 
depths at the radii of emission $r_{\rm {ej}}$ and re-intersection $r_{\rm{int}}$ to determine 
what fraction of ejecta get reabsorbed at $r_{\rm {ej}}$ rather than at $r_{\rm{int}}$. 
In general, this function depends on the direction in which the ejecta are launched, because
the probabilities are determined by slant path optical depths.
At low to moderate $\tau$'s, ejecta may execute more than one orbit before being 
absorbed at one or the other of these radii. 

\subsection{Ejecta Emission Rate}
\label{sub:rate}

The loss and gain integrals also contain an ejecta emission rate function
which depends on $r$ and $\tau$. As given by Eq. (9.6) of E18, CD90 found that a
good approximation to the $r$ and $\tau$ dependence of the local ejecta
mass emission rate per unit area\footnote{This function was determined for Oort
Cloud projectiles. It will probably differ somewhat for the EKB population (see
Sec. \ref{subsec:parameters}).} is

\begin{equation}
\label{equ:Rtau}
{\mathcal{R}}(r,\tau) = {\mathcal{R}}_\infty(r)\left[1 - e^{-\tau/\tau_0} + \left(\frac{\tau}{\tau_0}
\right)e^{-\tau/\tau_0}\right],
\end{equation} 

\noindent
where $\tau_0 \approx 0.28$. $\mathcal{R}_\infty$ is the ejecta 
mass emission rate per unit area as $\tau \rightarrow \infty$. Note that the
$\tau$-dependent function in Eq. (\ref{equ:Rtau}) peaks at $\tau = 2 \tau_0$
where $\mathcal{R} \approx$ 1.135 $\mathcal{R}_\infty$
As $\tau \rightarrow \infty$, $\mathcal{R}_\infty \approx$ 0.933 ${\mathcal{R}}(\tau=1)$.
In the limit of small $\tau$, as $\tau \rightarrow 0$, 

\begin{equation}
\label{equ:Rtausmall}
{\mathcal{R}}(r,\tau) \rightarrow 2{\mathcal{R}}_\infty(r) \tau/\tau_0.
\end{equation}

To an approximation good enough for our purposes,

\begin{equation}
\label{equ:Rinfty}
\begin{split}
{\mathcal{R}_\infty}(r) \approx \;<Y>\dot{\sigma}_{\rm{im}}(\tau \rightarrow \infty)
\approx \\
2<Y>\dot{\sigma}_\infty F_{\rm{g}}(r/r_0)^{-0.8},
\end{split}
\end{equation}

\noindent
where $r_0 = 1.8$ R$_{S}$ ($R_{\rm{S}} = 60,268$ km is Saturn's equatorial 
radius) and $\dot{\sigma}_\infty$ is the one-sided 
flat plate meteoroid 
influx in a heliocentric frame far from Saturn (for cometary projectiles) or in the 
Saturn reference frame far from Saturn (for the EKB projectiles).
The ``gravitational focusing factor" $F_{\rm g}$ is meant to characterize the 
focusing expected for an assumed population of meteoroid impactors. As 
discussed in Section 5.1 and E18, it is expected to be of order 3 for cometary
meteoroids and 30 for EKB meteoroids (see also Appendix B of CE98).
The $(r/r_0)^{-0.8}$ factor is adopted from CD90, based on a fit to
CD90 results, as an estimate for how the gravitational focusing depends 
on $r$ (see also Section 4.2 of CE98). 

\section{Mass Influx Models for Ballistic Transport and Mass Loading}
\label{sec:inflow}

Previous papers on BT and ML mostly concerned how structure in
the rings can be produced near edges (D89; D92; D96; E15; E18) 
or due to instabilities \citep{Dur95,Lat12,Lat14a,Lat14b}. 
This paper focuses instead on a more fundamental feature of BT and ML, 
namely that both mechanisms can drive significant mass
inflow. Inflow due to ML was discussed in D96, E15, and E18 in the
context of the lifetime of the current C ring.

\subsection{BT in a Quasi-Steady Uniform Ring}
\label{subsec:uniform}

We first address BT by considering the  
special case of a disk with constant $\Sigma(r)$ and $\tau(r)$. 
As demonstrated in D89, E15, and E18, this results in a steady-state 
behavior to a high degree of approximation away from boundaries, 
with deviations only to second order in $x$. Even the largest ejection 
speeds in the power-law distribution are expected to be only 100's m s$^{-1}$, leading
to $x \sim {\rm{few}} \cdot 10^{-3}$ or $x^2 \sim 10^{-6}$ to $10^{-5}$. In other 
words, it would take about $10^5$ to $10^6$ $t_{\rm{G}}$ to see large deviations 
from the steady state, a time comparable to or greater than a Solar System lifetime. 
Just as for a steady-state gaseous accretion disk, the uniform ring case is only an 
idealization and is significantly affected by the imposition of realistic boundary 
conditions.

We explore this analytically by making some approximations 
similar to those made by \citet{Lis84} and \citet[][herafter D95]{Dur95}. First 
suppose that there is only one ejecta speed, characterized by $x$, and that
all the ejecta are prograde. Suppose further that the re-intersection radii
of the prograde ejecta  are distributed uniformly from their
radius of origin $r_{\rm{ej}}$ to $r_{\rm{ej}} + \delta r_{\rm{max}}$, where
$\delta r_{\rm{max}}$ is the maximum ``throw distance" from 
$r_{\rm{ej}}$ that an ejectum can reach. To first order in $x$, D89 showed, using Keplerian orbits, 
that $\delta r_{\rm{max}} \approx 4xr_{\rm ej}$. This drastically simplifies the angle and
speed integrations in the $\Gamma$'s and $\Lambda$'s. 

The gain and loss integrals also contain relative probabilities of absorption 
at $r_{\rm{ej}}$ and $r_{\rm{int}}$. As discussed in D95 \citep[see also][]{Lat12},
a reasonable approximation to this probability function for a uniform ring is 

\begin{equation}
\label{equ:P}
P(\tau) = \frac{1 - e^{-\tau/\tau_p}}{1 - e^{-2\tau/\tau_p}},
\end{equation}

\noindent
as applied at $r_{\rm int}$, where $\tau_p \approx 0.5$ represents a typical slant path through the ring.
This allows $P$ to be taken outside the angle integration.

There are two limiting cases of Eq. (\ref{equ:P}). For $\tau \rightarrow 0$ the probabilities
are 1/2. An ejectum emitted at $r_{\rm{ej}}$ is just as likely to be reabsorbed at
$r_{\rm{ej}}$ or $r_{\rm{int}}$. For $\tau \rightarrow \infty$, on the other hand, the
probability of absorption at $r_{\rm{int}}$ is unity and the
probability of reabsorption at $r_{\rm{ej}}$ is zero. We will consider these two limiting cases, 
and then the results for a general $\tau$.

The expressions in Eqs. (\ref{equ:unif1}) to (\ref{equ:unif6}) below are cited by reference
to equations from appendices in D92 and E18. The reader who wishes to verify these 
results can use the linearized gain and loss integrals given, in slightly different notation, 
by Eqs. (22) to (30) of D95\footnote{A factor of $h_{\rm{c}}$ is erroneously missing
from the RHS's of the angular momentum loss and gains 
integrals in Eqs. (24) and (25) of D95. This omission
is a typo and does not affect any other results in D95.}.
Set $A$ and $B = 0$ in those equations and adopt $g(\cos\theta) = 1$ 
from $\cos\theta =$ 0 to 1 and $g(\cos\theta)=0$ otherwise. The $P_0$ in D95 is $P(\tau)$ 
given above, with 1/2 for very low $\tau$ and 1 for very high $\tau$. The zero-order terms
cancel when the gain and loss integrals are subtracted, and the first-order terms lead to the 
RHS's of Eqs. (\ref{equ:unif1}) to (\ref{equ:unif6}) below.

Insight into the physics
can be gained from understanding the $x\, {\rm cos}\,{\theta}$ terms in
the equations of D95.
The factor $1 - 8x\,{\rm cos}\,\theta$ in $\Gamma_{\rm m}$ comes from the
flat disk geometry of the rings, which causes the ring area elements for emitted ejecta
to be smaller than the ring area elements that they are transported to. Hence, for 
prograde ejecta, $\Lambda_{\rm m} < \Gamma_{\rm m}$ to order $x$ 
because ejecta emission annuli have less area 
than the absorption annuli. In other words, 
less mass is gained from inner ring regions than is lost to outer regions for
purely prograde ejecta, leading to a decrease in $\Sigma$ due to direct mass 
exchange. The difference in the angular momentum exchange integrals is that
there is an extra factor $1 - x{\rm cos}
\,\theta$ in $\Gamma_{\rm h}$ due to the
$x$ and ${\rm cos}\,\theta$ dependence of the ejecta angular momentum. So the 
radial mass influx is non-zero to order $x$. For a uniform ring, the divergence
of the radial flux exactly cancels the mass deposition to order $x$.

{\it Low Optical Depth.}
For a uniform ring, far from any boundaries, with these approximations, Appendix B of
D92 and Appendix A of E18 show that we get, with $\tau \rightarrow 0$, 
to first order in $x$, 

\begin{equation}
\label{equ:unif1}
\Gamma_{\rm{m}} - \Lambda_{\rm{m}} = -2x{\mathcal{R}}
\end{equation}

\noindent
and 

\begin{equation}
\label{equ:unif2}
v_{\rm{r}}^{\rm{ball}} = - xr{\mathcal{R}}/\Sigma.
\end{equation}

\noindent
If these expressions are put into Eq. (\ref{equ:mass}) with the mass loading term 
set to zero, for simplicity, we get that $\Sigma(r)$ is constant to first order in $x$.

{\it High Optical Depth.}
In the limit $\tau \rightarrow \infty$, these same expressions become, to first
order in $x$, 

\begin{equation}
\label{equ:unif3}
\Gamma_{\rm{m}} - \Lambda_{\rm{m}} = -4x{\mathcal{R}_\infty}
\end{equation}

\noindent
and 

\begin{equation}
\label{equ:unif4}
v_{\rm{r}}^{\rm{ball}} = - 2xr{\mathcal{R}_\infty}/\Sigma.
\end{equation}

{\it General Optical Depth.}
The general case has the $\tau$-dependent $P$ function.

\begin{equation}
\label{equ:unif5}
\Gamma_{\rm{m}} - \Lambda_{\rm{m}} = -4xP{\mathcal{R}}
\end{equation}

\noindent
and 

\begin{equation}
\label{equ:unif6}
v_{\rm{r}}^{\rm{ball}} = - 2xrP{\mathcal{R}}/\Sigma.
\end{equation}

\subsection{Inflow Rates for the Uniform Ring}
\label{subsec:inflowrates}

{If we multiply (\ref{equ:unif6}) on both sides by $\Sigma$, we get mass inflow in units}
of mass per unit length of circumference per unit time. So, multiplying by $2\pi r$ gives
the total mass inflow rate $\dot{M}_{\rm{bt}}$,

\begin{equation}
\label{equ:Mdotonex}
\dot{M}_{\rm{bt}} = 4\pi x r^2 P{\mathcal{R}},
\end{equation}

Strictly speaking, Eq. (\ref{equ:Mdotonex}) is valid only for a single
value of $x$, but, when there is an $x$-distribution, each $x$ will 
contribute to the influx according to Eq. (\ref{equ:Mdotonex}) but with appropriate
weighting. Let's assume that $x$ obeys an $x^{-3}$ power law from
a lower limit $x_b$ up to an upper value $x_t$ much like what is assumed
in D92 and E15. Then, for $x_t >> x_b$, the average $x$ weighted by
the power law is $2x_b$, and we get the case for general $\tau$ and an
$n = 3$ power law as

\begin{equation}
\label{equ:Mdotmultix}
\dot{M}_{\rm{bt}}(\tau) = 8\pi x_b r^2 P{\mathcal{R}}.
\end{equation}

\noindent
Inserting the appropriate limiting cases for $P$ and $\mathcal{R}$, we
get, for $\tau \rightarrow 0$,

\begin{equation}
\label{equ:Mdottausmall}
\dot{M}_{\rm{bt}}({\rm{low}}) = 16\pi x_b r^2 (\tau/\tau_0) <Y>\dot{\sigma}_\infty F_{\rm{g}}(r/r_0)^{-0.8}
\end{equation}

\noindent
and, for $\tau \rightarrow \infty$, 

\begin{equation}
\label{equ:Mdottaularge}
\dot{M}_{\rm{bt}}(\infty) = 16\pi x_b r^2 <Y>\dot{\sigma}_\infty F_{\rm{g}}(r/r_0)^{-0.8}
\end{equation}

While these mass inflow rates $\rightarrow 0$ as $\tau \rightarrow 0$, 
it is important to realize that the BT radial inward drift speed $v_{\rm{r}}^{\rm{ball}}$
does not $\rightarrow 0$ as $\tau \rightarrow 0$.
The terms involving the $\Gamma$'s and $\Lambda$'s
in the second bracket on the RHS of  Eq. (\ref{equ:vrball}) are proportional to $\tau$
at low $\tau$ and hence, at constant opacity, to $\Sigma$, while the first factor on the RHS 
contains a $1/\Sigma$. So, as $\tau \rightarrow 0$, the $\tau$ and $\Sigma$-dependences
cancel out in the expression for $v_{\rm{r}}^{\rm{ball}}$. 
This same effect applies to $v^{\rm load}_{\rm r}$.

\begin{figure}
 \resizebox{\linewidth}{!}{%
 \includegraphics{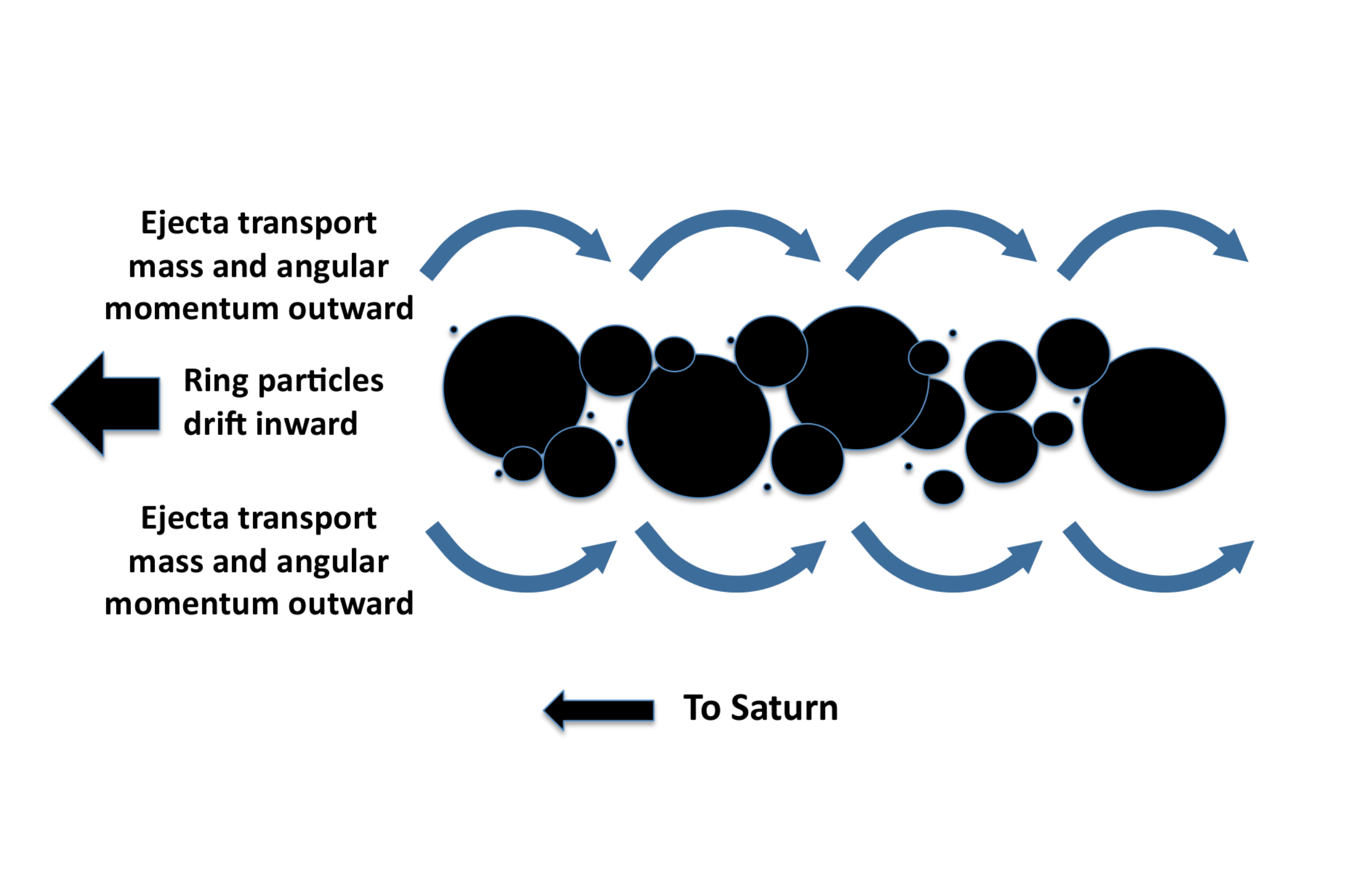}}
 \vspace{-0.5in}
\caption{Cartoon illustrating mass and angular momentum 
flows for the steady-state uniform ring. Angular momentum and mass are carried
outward by the prograde impact ejecta. The result is that the local specific 
angular momentum of the ring particles decreases and they drift 
inward {\it en masse}. The combined effect of the direct mass and angular
momentum transport by the ejecta and the ring particle drift is that ring 
surface density remains constant to first order in the small ejecta speed
parameter $x$.}
\label{fig:cartoon}
\vspace{-0.1in}
\end{figure}

\subsection{The Uniform Ring as Analog to a Steady-State Accretion Disk}
\label{subsec:accretiondisk}

The uniform ring resembles a steady-state viscous accretion disk \citep[see, e.g.,][]{Har09}. 
In both cases, angular momentum
is transported outward while mass flows inward without changing the local surface density.
In the BT uniform ring, as illustrated in {\bf Figure \ref{fig:cartoon}}, 
ejecta flying mostly outward above the ring plane carry mass and angular momentum
outward and cause changes in the local surface density and specific angular momentum.
Because angular momentum is transported inexorably outward by the prograde ejecta, 
the specific angular momentum of the ensemble of ring particles in the ring plane 
decreases  so that it drifts inward {\it en masse}. The combined effect is that $\Sigma$ remains
constant locally in a uniform ring system to a high degree of approximation.
The entire ensemble of ring particles consequently moves inward
like a conveyor belt because angular momentum is being transported outward.
The analogy is not perfect of course. 
In a viscous accretion disk, it is viscous stresses that carry angular momentum
outward, and the steady-state surface density profile depends on the viscosity law. 

Realistically, there will be inner and outer boundaries where the BT uniform ring
approximation will not apply. Still, the inner boundary region, whatever it is -- 
perhaps the atmosphere of the planet or a region where other mechanisms operate 
effectively to remove mass -- must accommodate the mass inflow. Similarly, the
angular momentum carried outward by the ejecta must accumulate somewhere or
be removed by other processes. Similar boundary condition issues arise for a
viscous accretion disk.

\subsection{Mass Loading}
\label{subsec:massload}

The direct deposition of mass and angular momentum by meteoroids
sets a lower bound on the radially inward drift speed 
and mass inflow rate that is independent of uncertain parameters 
like $x_b$ and $<Y>$. 
Fig. 1 of D96, which is based on CD90, demonstrates the non-zero inward 
radial drift speed as $\tau \rightarrow 0$ due to mass loading and meteoroid torque.
It also shows that the meteoroid torque effect on radial drift at low $\tau$ is only about 1/4 that of mass 
loading alone. So, for a first estimate of inflow rate and speed due to 
meteoroid deposition, we consider only the direct deposition of mass 
and not the meteoroid torque. In other words, we
ignore the $\dot{J}_{\rm{im}}$ term in Eq. (\ref{equ:vrload}).

D96 made analytic fits, given by their Eqs. (7) and (8), 
to the numerical results of CD90 for cometary 
dust {to get $v^{\rm load}_{\rm r}$} 
{as a function of $\tau$ and $r$. Here we adopt the D96 work but
use simpler analytic forms that still reflect what 
we know about the expected dependences on variables $\Sigma$, $\tau$ and $r$ 
and on the parameter $F_{\rm g}$. We replace the 
$\tau^{-1}$  with $\Sigma^{-1}$ by calling out the assumption 
in D96 that the ring opacity has a constant value of about 0.01 cm$^2$ g$^{-1}$. 
Also, the $\tau$-dependence of $v^{\rm{load}}_{\rm{r}}$ should  
be the same as the $\tau$-dependence of $\dot{\sigma}_{\rm{im}}$ given 
by Eq. (2) in E15 (see also Eq. [1] of D96), also based on fits to 
CD90 computations. We have simplified the D96 $\tau$-dependence by 
setting the $k$ and $k_{ml}$ parameters that appear in the D96 equations equal to one.
The linear $r$-dependence in Eq. (8) of D96 is actually just a crude approximation
to the power-law $r$-dependence of the focussing. Here we more accurately reflect 
that $r$-dependence of focussing explicitly.}
 
{The result of all these modifications is a more generally 
applicable and versatile expression, which 
represents the CD90 computations as accurately as do Eqs. (7) and (8) of D96. 
The revised equation is}

\begin{equation}
\label{equ:massload}
v^{\rm{load}}_{\rm{r}}(\rm{CD}) = -  {\it s}_{\rm{load}} 
({\it r} / {\rm{R}_{\rm{S}}})^{0.2}(1 - e^{-\tau/\tau_{\rm{load}}})/\Sigma, 
\end{equation}

\noindent
where $\tau_{\rm{load}} = 0.52$,  

\begin{equation}
\label{equ:cload}
s_{\rm{load}} =  7.2\cdot10^{-6} (F_{\rm g}/3) \; \rm{g}\: \rm{cm}^{-1}\; s^{-1},
\end{equation}

\noindent
and the ``CD" refers to the CD90 paper, in which Oort Cloud
meteoroid projectiles were assumed, 
{giving $F_{\rm g} \approx 3$. 
The factor of $(r / {\rm{R}_{\rm{S}}})^{0.2}$ in Eq. (\ref{equ:cload}) 
properly includes the $(r / r_0)^{-0.8}$ factor from Eq. (\ref{equ:Rinfty}) due to 
the radial dependence of the gravitational focusing. Note that, because $\Sigma$
is proportional to $\tau$ for constant opacity, $v^{\rm{load}}_{\rm{r}}$ has its 
largest magnitude as $\tau \rightarrow 0$, with important implications
for ring lifetimes (see Section \ref{subsec:load}).}

\section{Application to Saturn's Rings}
\label{sec:rings}

\subsection{Input Parameters}
\label{subsec:parameters}

The RHS's of Eqs. (\ref{equ:Mdotmultix}), (\ref{equ:Mdottausmall}), 
and (\ref{equ:Mdottaularge}) depend on 
location in the rings $r$, the optical depth $\tau$ at $r$, the typical impact yield $<Y>$,
the $x_b$ characterizing the low end of the power-law ejecta speed distribution,
the meteoroid influx at large distance from Saturn $\dot{\sigma}_\infty$, and the 
strength of the gravitational focusing $F_{\rm{g}}$, which in turn depends on the meteoroid
source population. It is worth noting that the RHS's depend only on $\tau$, not on
$\Sigma$. Parametrically, what matters for mass inflow is the efficiency of meteoroid 
absorption, through $\tau$, not the particle size distribution, particle porosity,
ring texture, or ring opacity. This inflow rate 
is also independent of viscosity.

On the other hand,
the factor $<Y>$ is likely to depend on ring particle properties, but exactly how is 
not well understood. Let $Y$ without brackets denote the {nominal yield} for cratering 
impacts with normal incidence at a speed of 14.4 km s$^{-1}$.
Previous models of the rings' structural evolution showed that 
$Y \gtrsim 10^6$ was required to maintain the sharpness of the inner B ring 
edge and to form the linear ramp that connects it to the C ring (D92; E15). These 
calculations, however, employed the old cometary flux. The most recent models using the 
more focused EKB flux adopted in this paper only require $Y \gtrsim 10^5$ 
\citep[][E18]{Est16,Est17}. Moreover, analyses of ring pollution ages (E15; E18) 
seem roughly consistent with $Y \gtrsim 10^4$ to $10^5$. Thus we adopt 
$<Y>\; =10^5$ as our fiducial value for this work, and this is equivalent
to choosing $Y$ to be about the same value. 

In BT simulations, we have typically adopted 
$x_b = 10^{-4}$. This parameter is not well constrained, but is roughly consistent with 
matching structures near the inner edge of Saturn's B ring (E15). In terms of the nature 
of impacts on the rings, both $x_b$ and especially $Y$ are uncertain by up to an order of magnitude,
possibly more, and could vary with location depending on ring particle properties like shape and porosity. 

The CDA observations \citep{Kem17,Alt18,Kem23} indicate that dust-sized meteoroid impacts are
dominated by an EKB source and that the value {of the micrometeoroid flux at infinity is} 
$\dot{\sigma}_\infty = \dot{\sigma}_\infty{\rm{(CDA)}} = 2.2\cdot 10^{-17}$ g cm$^{-2}$ s$^{-1}$. 
{This value} 
{is lower by a factor of two compared to} 
earlier estimates for the influx 
of small meteoroids at Saturn of $\dot{\sigma}_\infty = \dot{\sigma}_\infty({\rm{CD}}) = 4.5\cdot 10^{-17}$ g 
cm$^{-2}$ s$^{-1}$ (Gr\"un \etal, 1985; 
CD90; CE98; Sec. \ref{subsec:pollution}),
{where again ``CD'' refers to the value adopted in CD90, and} 
is generally consistent with measurements of
the interplanetary meteoroid flux by the New Horizons Student Dust Detector 
\citep{Pop16,Pop19a,Pop19b}. For simplicity, in this paper, we assume 
the influx at Saturn is constant over relevant time scales.

The CDA measurements also give information about the velocity distribution
of the meteoroids as they enter the Saturn system. We have not yet performed a 
detailed re-computation of the focusing and aberration for EKB meteoroids along the 
lines of CD90. However, E15 and E18 estimated that the focusing factor 
$F_{\rm{g}}$ is $\sim 30$ for these objects rather than the three or four assumed by CD90 and CE98 
for cometary dust, because the EKB particles enter Saturn's Hill sphere with smaller 
speeds relative to Saturn (see CD90; CD98{; and the discussion in Section \ref{subsec:pollution}).}
The exact functional form of $\mathcal{R}(\tau)$, 
based here on CD90, is likely to change due to a different degree of aberration
for EKB particles, but we suspect this probably only affects results by factors of order unity.

The values of $<Y>$ used here are rather large but, as explained in E18, 
are adopted based on pollution models and fits to ring structure, 
especially the outer C ring and inner B ring regions. As noted above, using 
cometary meteoroids, D92 and D96 argued for $<Y> \sim 10^6$ 
to match the edge/ramp structure. In their laboratory experiments 
for impacts onto ice and ice/silicate targets, {\citet{KG01a}} found 
it was difficult to justify $Y$ being as large as $10^6$ and suggested that other 
highly uncertain parameters in ballistic transport modelling might {have to be}
adjusted to bring the required yield down. 

{The CDA results already move in the right direction.} The increase in $F_{\rm g}$ 
due to the EKB nature of the impactors has now reduced the estimated $Y$ for 
structural fits by {about an} 
{order of magnitude (E15;E18)}. 
With remaining uncertainties in viscosity,
in the upper and lower limits of the ejecta velocity power law, and in the exact value
of the power law index $n$,
plus the absence of a proper and complete treatment 
of meteoroid focusing and aberration, the value of the structurally required $Y$ 
could drop further and be brought more in line with laboratory results. 

{Another concern raised by \citet{KG01a} is illustrated by their Eq. (7) and Table III.
Using parameters appropriate for icy projectiles and targets, their experimentally 
determined yields for a normal impact at 14 km s$^{-1}$ can be approximately fit by 
$1.2\cdot 10^5(r_{\rm i}/1 {\rm cm})^{0.69}$, where 
$r_{\rm i}$ is the radius of the impactor. The particles detected
by CDA are generally 
less than 100 $\mu$m\footnote{{One grain considerably larger than 100 $\mu$m was detected which would increase the magnitude of $\dot{\sigma}_\infty$, but was not included in the estimate of the flux for reasons 
discussed in \citet{Kem23}. These authors also discuss the probabilities of non-detection of larger grains due to insufficient detector exposure time.}}, implying yields from this fit of less than about 
5$\cdot$10$^3$. 
Here we 
adopt values of $<Y>$ compatible with the inner edge
results of BT evolutions. 
It should be noted that it is actually the combination $\dot{\sigma}_\infty<Y>/\nu$ 
that matters for matching inner edge structures (E15; E18). Any deficiency in the actual
$<Y>$ could be compensated by an increase in $\dot{\sigma}_\infty/\nu$. We should also 
keep in mind that, despite best efforts, the laboratory impact results 
may not apply well to ring particles with regoliths and with complex substructure 
that are in free-fall at cryogenic temperatures.}

Another concern is how well our inflow models may apply to parts of the
B ring where observations suggest high porosity of the ring particles \cite[e.g.,][]{Ref15,Zha17a},
and similarly for observations in the A and C rings \citep{Por08, Mor16,Zha17b}.
High porosity may imply lower yields so that possibly the inflow rate in the central 
part of the B ring is actually lower. Pollution models and BT structural arguments do not 
constrain $Y$ in the middle B ring. However, this
would not affect our explanation of how mass flows from the inner B ring and 
through the C ring toward the planet, which is the main focus of this paper. 
On the other hand, the yield in regions of porous particles may well be just as large as elsewhere 
if the typical compacted components of the porous particle aggregates are 
{significantly larger than the size of the impacting meteoroids. 
Then impacts by meteoroids could produce yields corresponding 
to the individual components of the aggregate, not the porous aggregate as a whole.}

{So, with all these caveats, we here consider a range of $<Y>$ from $10^4$ to $10^5$.} 
 
 \begin{figure}
 \resizebox{\linewidth}{!}{%
 \includegraphics[width=2in]{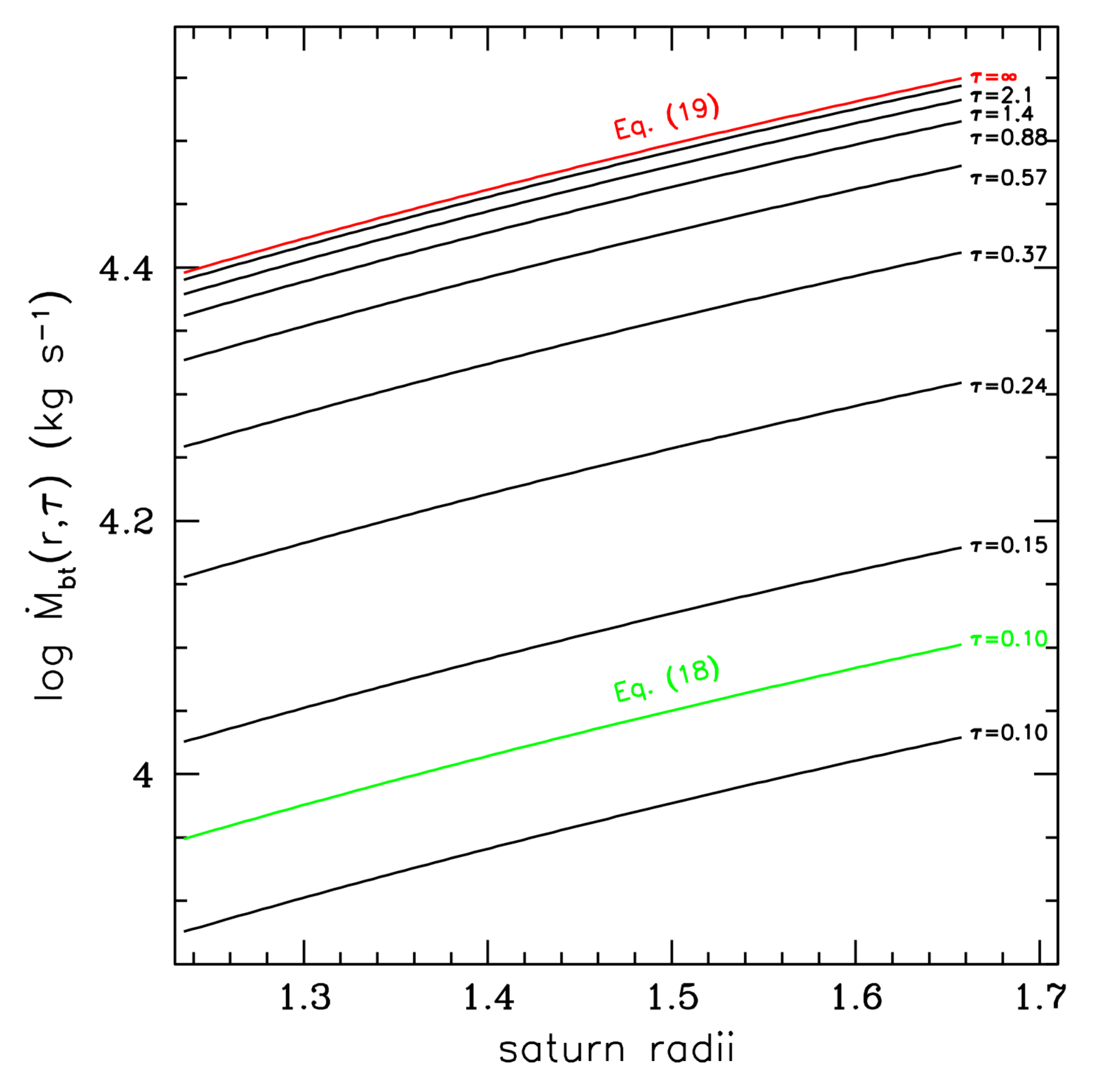}}
 \vspace{-0.3in}
\caption{Log of the mass inflow rates (Eq. [\ref{equ:Mdotmultix}]) in kg s$^{-1}$ 
due to BT as a function of distance from Saturn for a range of optical depths (solid black curves). 
Also plotted are the $\tau \rightarrow 0$ (Eq. [\ref{equ:Mdottausmall}]) and $\tau \rightarrow \infty$ 
(Eq. [\ref{equ:Mdottaularge}]) limits in green and red, respectively. Equation
(\ref{equ:Mdottaularge}) is a reasonable choice for $\tau \gtrsim 2$; however, Eq. (\ref{equ:Mdottausmall}) 
overshoots the correct value for $\tau = 0.1$ (see Fig. \ref{fig:Mdotlimits}). 
For reference, C ring optical depths lie in the range of $\tau \sim 0.05-0.5$, while the B ring has
$\tau > 1$. The inner B ring edge lies at 1.52 R$_S$. }
\label{fig:mdotin}
\vspace{-0.1in}
\end{figure}
 
\subsection{Mass Inflow due to Ballistic Transport}
\label{subsec:Mdotnumbers}

Altogether, Eqs. (\ref{equ:Mdottausmall}) and (\ref{equ:Mdottaularge}) then become

\begin{equation}
\label{Mdotnumbertausmall}
\begin{split}
\dot{M}_{\rm{bt}}({\rm{low}}) = 1.1\cdot10^4\; {\rm{kg}\; \rm{s}^{-1}} 
\left( \frac{x_b}{10^{-4}} \right) \left(\frac{<Y>}{10^5} \right) \times \\
\left( \frac{\dot{\sigma}_{\infty}}{\dot{\sigma}_{\infty}{\rm{(CDA)}}} \right)
\left(\frac{F_{\rm{g}}}{30} \right) \left( \frac{r}{1.5 {\rm{R}}_S}\right)^{1.2}
\left( \frac{\tau}{0.1}\right),
\end{split}
\end{equation}

\noindent
and

\begin{equation}
\label{Mdotnumbertaularge}
\begin{split}
\dot{M}_{\rm{bt}}(\infty) = 3.2\cdot10^4\; {\rm{kg}\; \rm{s}^{-1}} 
\left( \frac{x_b}{10^{-4}} \right) \left(\frac{<Y>}{10^5} \right) \times \\
\left( \frac{\dot{\sigma}_{\infty}}{\dot{\sigma}_{\infty}{\rm{(CDA)}}} \right) \left(\frac{F_{\rm{g}}}{30} \right)
\left( \frac{r}{1.5 {\rm{R}}_S}\right)^{1.2}.
\end{split}
\end{equation}

\noindent
Although, given the uncertainties in the parameters,  
the true inflow values could be up to an order of magnitude higher or lower, 
these mass inflow rates capture the 
principal point of this paper, namely that, for reasonable parameters, 
mass inflow driven by BT alone is comparable to the mass inflow
into Saturn measured by Cassini during its final orbits, 
$\sim$ few $\cdot\, 10^3$ to a few $\cdot\, 10^4$ kg s$^{-1}$. 
The high optical depth result is 
appropriate for the B ring, and the low optical depth result
is appropriate, in an average sense, for the C ring. 

\begin{figure}
 \resizebox{\linewidth}{!}{%
 \includegraphics{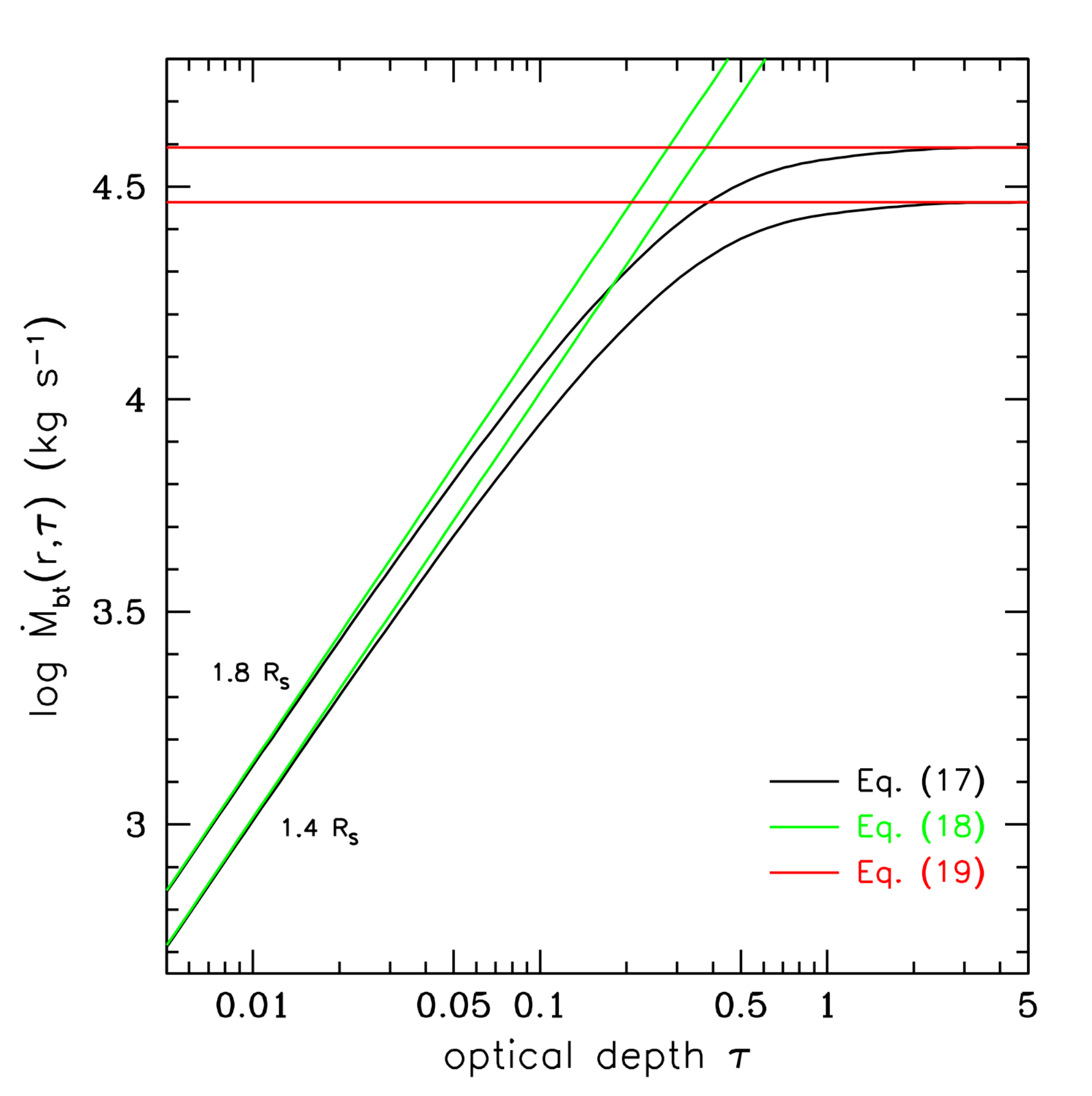}}
 \vspace{-0.3in}
\caption{The approximations $\tau \rightarrow 0$ (Eq. 
[\ref{equ:Mdottausmall}], green lines) and $\tau \rightarrow \infty$ (Eq. [\ref{equ:Mdottaularge}]
red lines) for BT inflow compared with the general $\tau$ case (Eq. [\ref{equ:Mdotmultix}], black curves)
plotted for two values of $r$, characteristic of the C ring and B ring. The high-$\tau$ limit is valid 
for $\tau \gtrsim 2$, while use of Eq. (\ref{equ:Mdottausmall}) should be
restricted to $\tau \lesssim 0.05$.}
\label{fig:Mdotlimits}
\vspace{-0.1in}
\end{figure}

More generally, {\bf Figure \ref{fig:mdotin}} shows Eq. (\ref{equ:Mdotmultix}) computed 
for a variety of $\tau$'s as a function of $r$ for the same nominal
parameter choices used in Eqs. (\ref{Mdotnumbertausmall}) and 
(\ref{Mdotnumbertaularge}). Note that the inflow rate asymptotes to a finite
value as $\tau \rightarrow \infty$. The result is that we
expect a rather constant high value of $\dot{M}_{\rm bt}$ for
a region where optical depths are generally greater than unity, like
the B ring. {\bf Figure \ref{fig:Mdotlimits}} shows how well the low-$\tau$ 
and high-$\tau$ limiting expressions match the inflow results for general $\tau$.

As a supplemental result, which will aid in later discussion, 
the gross erosion time in the inner B ring for the same parametrization is

\begin{equation}
\label{equ:teegee2}
t_{\rm{G}} \approx 10^4\; {\rm{yrs}}
\left(\frac{\Sigma}{52\; {\rm{g}}\; {\rm{cm}}^{-2}} \right) 
\left(\frac{10^5}{<Y>} \right)
\left(\frac{\dot{\sigma}_{\infty}{\rm{(CDA)}}}{\dot{\sigma}_{\infty}} \right).
\end{equation}

\noindent
{For this calculation, we have used the derived estimate for the B ring mass of $\sim 0.23$ Mimas masses
\citep{Ies19} spread out over the radial extent of the B ring from 92,000 to 117,580 km to obtain an average
surface density of 52 g cm$^{-2}$.}

\subsection{Mass Inflow due to Mass Loading}
\label{subsec:load}

Eq. (\ref{equ:massload}) gives the mass loading influx rate
for cometary dust. For EKB impacting meteoroids, we make the 
approximation that the gravitational focusing should be about a factor of ten higher (E18). 
{Adopting $F_{\rm g} \approx 30$ instead of 3 for EKB impactors, as well as a correction factor
to account for the CDA flux $\dot{\sigma}_\infty{\rm{(CDA)}}/\dot{\sigma}_\infty{\rm{(CD)}}$,
Eqs. (\ref{equ:massload}) and  (\ref{equ:cload}) become} 

\begin{equation}
\label{equ:vloadEKB}
\begin{split}
\Sigma v^{\rm{load}}_{\rm{r}}({\rm{EKB}}) =
- 3.5\cdot10^{-5}\; {\rm{g}}\; {\rm{cm}}^{-1}\; {\rm{s}}^{-1} (F_{\rm g}/30) \times \\
({\it r} / {\rm{R}}_S)^{0.2}(1 - e^{-\tau/\tau_{\rm{load}}}).
\end{split}
\end{equation}

\noindent
Multiplying by $-2\pi r$, gives the mass inflow rate driven by mass loading
for EKB impactors

\begin{equation}
\label{equ:MdotloadEKB}
\begin{split}
\dot{M}_{\rm{load}} = 
2.2\cdot10^{3}\; {\rm{kg}\; s^{-1}}\; (1 - e^{-\tau/\tau_{\rm{load}}}) \times \\
\left(\frac{\dot{\sigma}_{\infty}}{\dot{\sigma}_{\infty}({\rm{CDA}})} \right) \left(\frac{F_{\rm{g}}}{30}\right)
\left(\frac{r}{1.5 {\rm{R}}_S}\right)^{1.2}.
\end{split}
\end{equation}

\noindent
Eq. (\ref{equ:MdotloadEKB}) assumes zero meteoroid torque. 
It is likely that the meteoroid torque is not so negligible for
EKB impactors, due to their greater aberration. This could increase
the inflow rate noticeably and will be the subject of future research.

\begin{figure}
 \resizebox{\linewidth}{!}{%
 \includegraphics{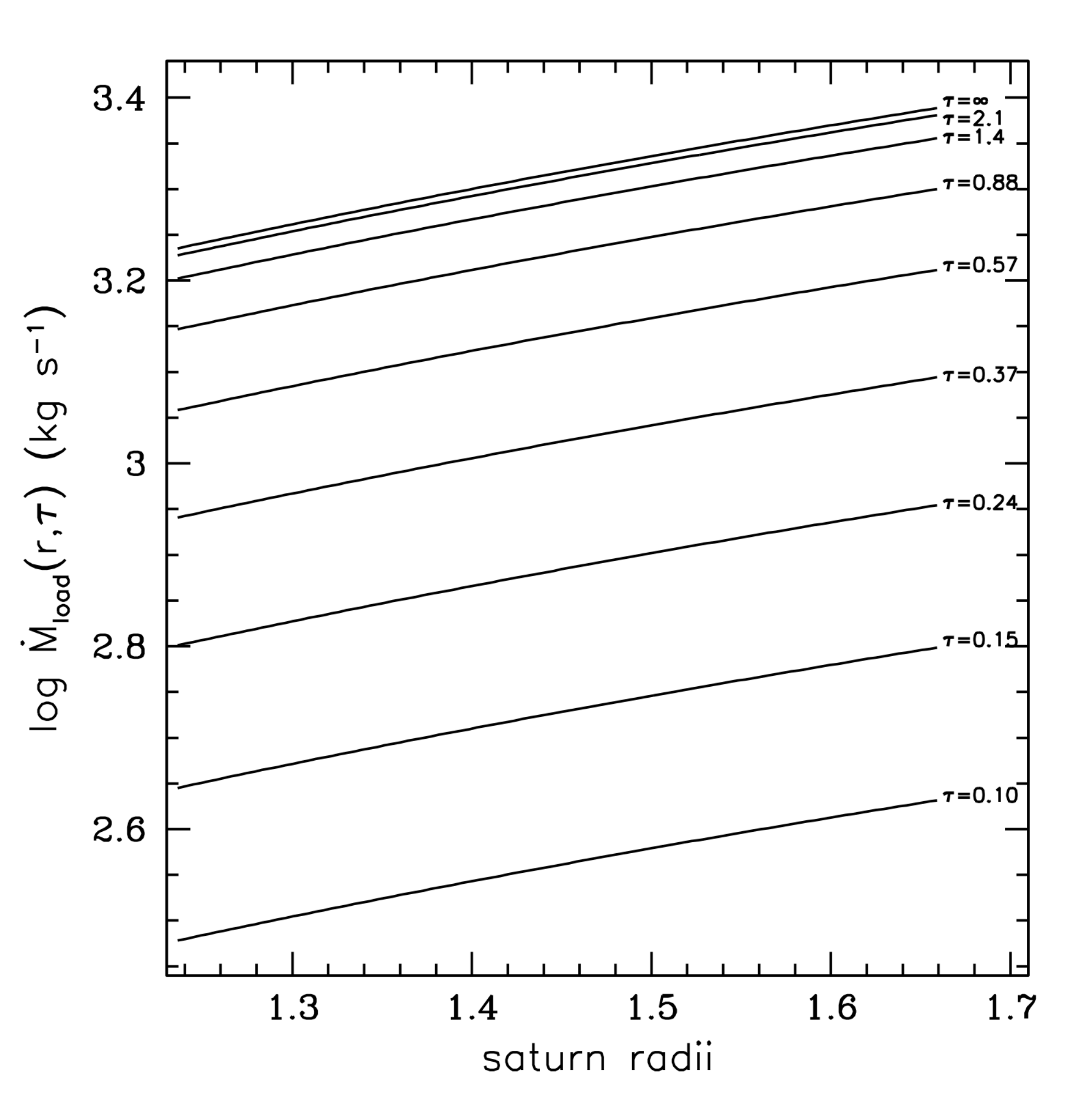}}
 \vspace{-0.3in}
\caption{Mass inflow rates due to mass loading as a function of $r$ for
various $\tau$-values using Eq. (\ref{equ:MdotloadEKB}) evaluated for the 
nominal choices of parameters.}
\label{fig:mdotload}
\vspace{-0.1in}
\end{figure}

{\bf Figure \ref{fig:mdotload}} shows the result of evaluating 
Eq. (\ref{equ:MdotloadEKB}) for a variety of $\tau$'s as a function of $r$.
Comparing this with Fig. \ref{fig:mdotin} reveals that
$\dot{M}_{\rm{load}}$ is roughly an order of magnitude smaller than $\dot{M}_{\rm{bt}}$,
{and} is, by itself, 
{slightly smaller than} the lower estimate of the mass inflow into Saturn
observed by INMS. If we reduce our choice of  $<Y>$ from $10^5$ to $10^4$, the inflow
rates for BT and mass loading are comparable in magnitude, {and in tandem are consistent with
the lower bound of the mass inflow}. 
Overall, the total mass inflow rate $\dot{M}_{\rm{total}} =  \dot{M}_{\rm{bt}}
+ \dot{M}_{\rm{load}}$ is quite large for $\tau > 0.1$ and $<Y>\; = 10^5$ 
and is consistent with the INMS upper estimate for the mass inflow. 

Note that the $r$-dependence for both effects is the 
same and is due to the combined effects of thin disk geometry and gravitational 
focusing. While the BT inflow depends on the uncertain parameters $x_b$ and
$<Y>$, which characterize the lowest speeds in the power-law ejecta speed 
distribution and the impact yield, respectively, mass loading depends only 
on the influx rate of meteoroids and their degree of focusing. 
With BT added, the mass influxes are larger
by an order of magnitude for $<Y>\; =  10^5$. We remind the reader that this value
of $<Y>$ is adopted here on the basis of BT structural arguments 
(Sec. \ref{subsec:parameters}).

\begin{figure}
 \resizebox{\linewidth}{!}{%
 \includegraphics{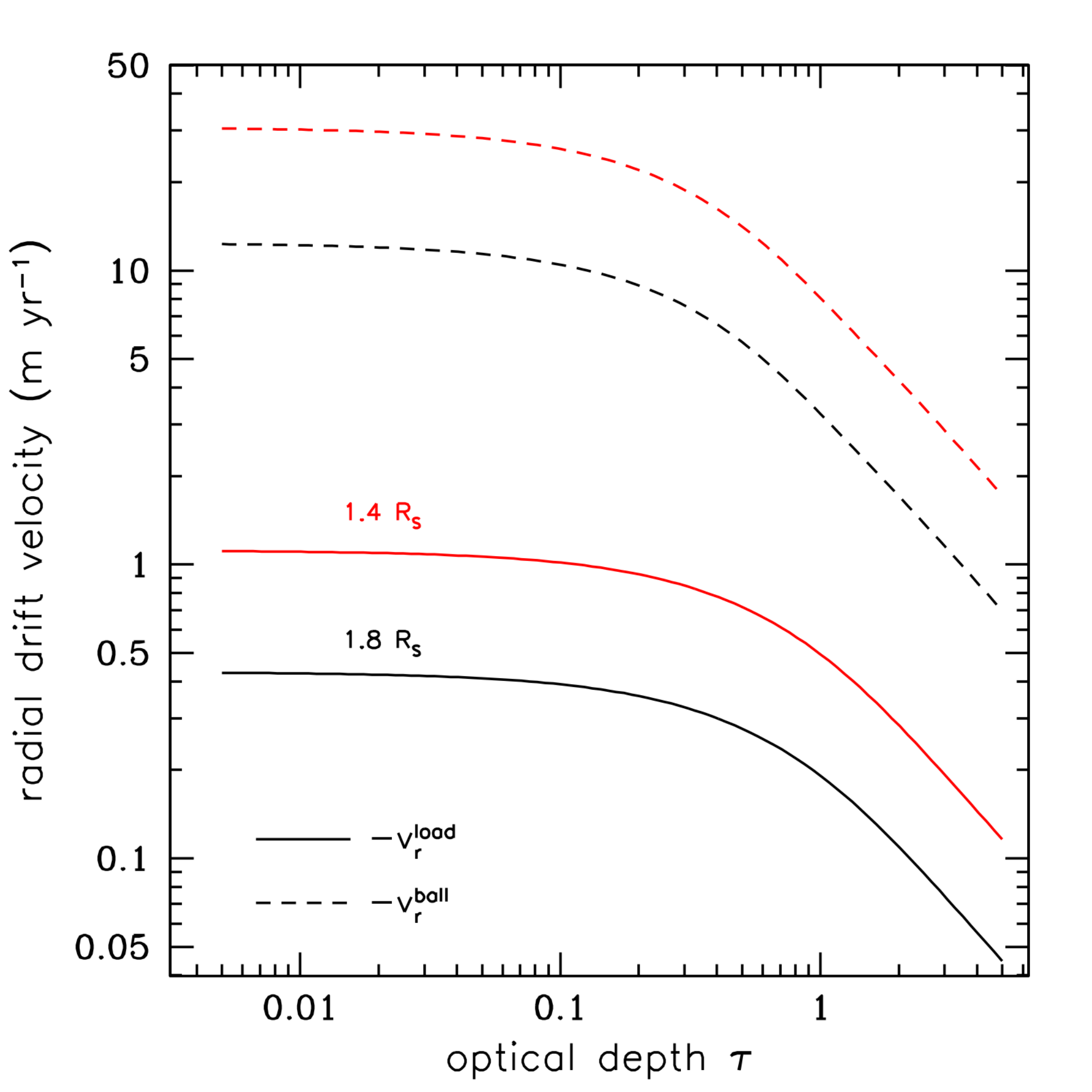}}
 \vspace{-0.3in}
\caption{Radial drift velocities due to mass loading (solid curves) and
BT (dashed curves) as a function of optical depth for characteristic locations in the C ring (red)
and B ring (black). For these curves, we have assumed opacities in the C ring and
B ring of $0.05$ and $0.013$ cm$^2$ g$^{-1}$, respectively. For both mass
loading and BT, the inward radial velocities asymptote to large values as optical
depth decreases, with implications for C ring formation.}
\label{fig:vdrift}
\vspace{-0.1in}
\end{figure}

In {\bf Figure \ref{fig:vdrift}}, we show the inward radial drift velocities due to
mass loading alone (solid curves) and due to BT alone (dashed curves)  
for characteristic locations in the C and B rings at 1.4 R$_S$ 
(red curves) and 1.8 R$_S$ (black curves), respectively, as a function of optical
depth. In generating these curves, we use Eq. (\ref{equ:vloadEKB}) 
for mass loading  and Eq. (\ref{equ:unif6}) integrated over the standard $x$-distribution
for BT. We also assume that $\Sigma = \tau/\kappa$ and use an opacity $\kappa = 0.05$
cm$^2$ g$^{-1}$ for the C ring and $0.013$ cm$^2$ g$^{-1}$ for the B ring 
{\citep[see, e.g., CE98;][]{Bai11,Hed11,Zha17a,Zha17b}.} 
For fixed opacity, the radial drift speeds increase 
to a nonzero asymptotic value as optical depth and surface density decrease, even though 
the mass inflow rate goes to zero. 

Based on Fig. \ref{fig:vdrift}, we expect  the low-$\tau$ base of the inner edge of an 
initially high-$\tau$ ring to spread inwards more quickly than the high-$\tau$ 
bulk of the ring.  Such sharp high-$\tau$ inner edges do occur in global viscous
evolutions of rings like those of \citet{Sal10} and \citet{CC12}.
The implication is that the C ring may have been
created from the original B ring inner edge due to BT or even to mass loading alone {\citep[see][]{ED23}}.

\begin{figure}
 \resizebox{\linewidth}{!}{%
 \includegraphics{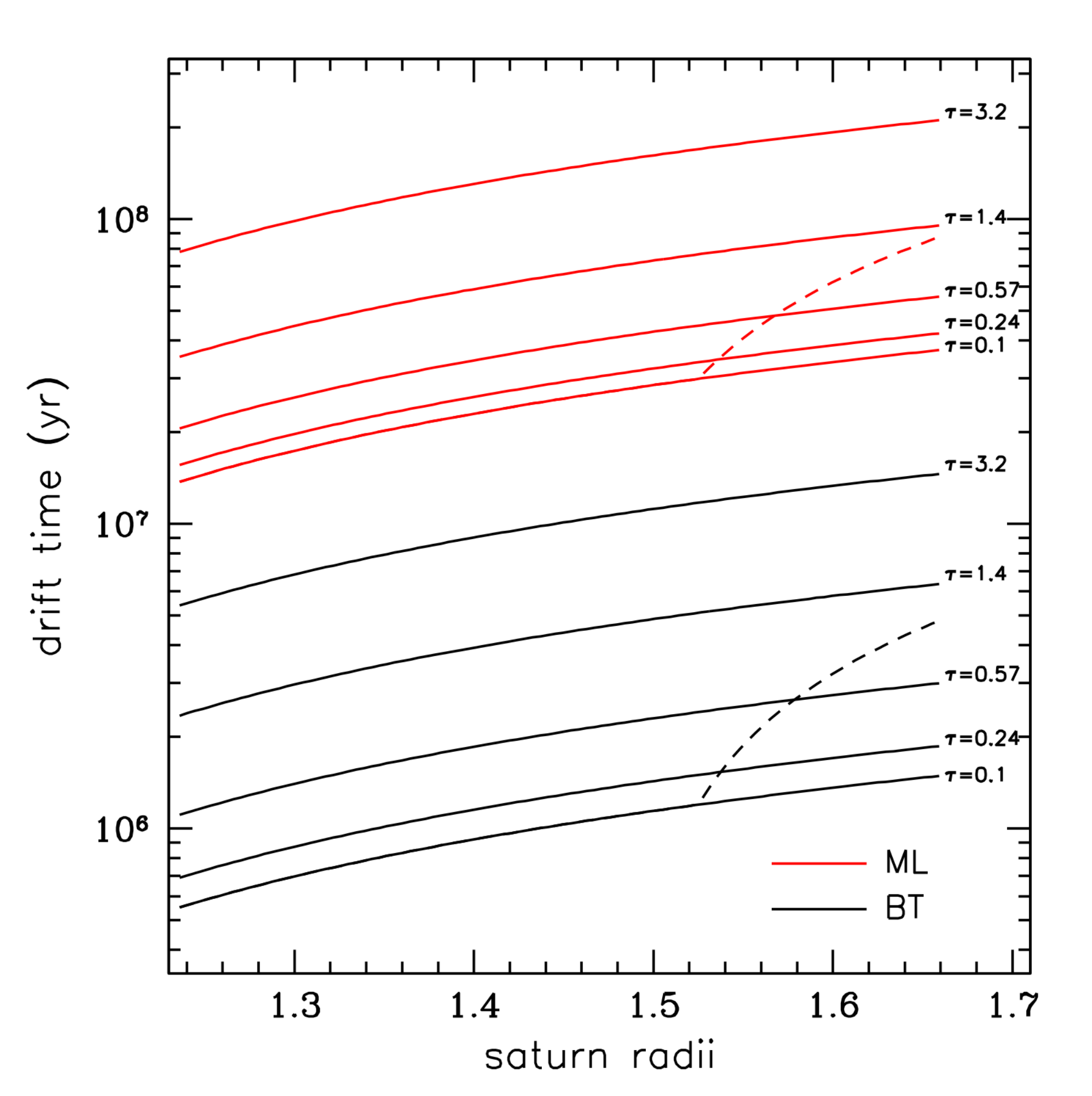}}
 \vspace{-0.3in}
\caption{Integrated time $t_{\rm{loss}}$ for ring material to fall into Saturn 
due to mass loading (red curves) and to BT (black curves) from radial locations in the C and inner B ring. 
Several values of optical depth are shown. The dashed curves are for a case in which we model the current rings
as a step function at $r = 1.525$ R$_S$ (the location of the inner B ring edge), where the optical 
depth and opacity are $\tau = 0.1$ and $\kappa = 0.05$ cm$^2$ g$^{-1}$ in the C ring and $\tau = 1$ 
and $\kappa = 0.013$ cm$^2$ g$^{-1}$ in the B ring.}
\label{fig:drift}
\vspace{-0.1in}
\end{figure}

These radial drift speeds can be used to estimate a time scale $t_{\rm{loss}}$ for ring material 
located at some radial distance $r$ to fall into Saturn. {\bf Figure \ref{fig:drift}} shows 
this time scale for mass loading (red curves) and BT (black curves) for several $\tau$. To generate 
these curves, we have again used Eq. (\ref{equ:vloadEKB}) and Eq. (\ref{equ:unif6}) integrated over $x$
to obtain inward drift speeds. We then integrate the drift speeds  over $r$  assuming constant optical 
depth and surface density and using the  C ring opacity in the previous paragraph. 
Drift times are much shorter for low $\tau$ and consistent with 
previous estimates that the C ring could be lost (or ``recycled'') in $\sim 10^7$ years by mass loading
alone. These times are much shorter for BT, but would be comparable to mass loading for $<Y>\; = 10^4$. We 
also see that $t_{\rm{loss}}$ becomes longer with increasing optical depth (or $\Sigma$), so
that, in an initially massive ring evolving due to strong viscosity, the effects of mass loading and  
ballistic transport are probably overwhelmed by viscosity until the ring mass decreases 
significantly.  

The dashed curves in Fig. \ref{fig:drift} are for a case in which we model the C ring and
inner B ring as a step function with $\Sigma = 2$ g cm$^{-2}$ for $r < 1.525$ R$_S$, and 
{$\Sigma = 52$ g cm$^{-2}$ for $r > 1.525$ R$_S$ (Sec. \ref{subsec:Mdotnumbers})}, representative of current values. 
For the B ring, this corresponds to taking $\tau \sim 1$.
{For the dashed curves loss time estimates, $\tau$ and $\Sigma$ 
at a given $r$ are kept constant with time, 
but the co-moving values change as the mass elements cross the inner B ring edge.}  
Note that the residence time for a mass element
in the inner B ring is quite short -- less than a few million years for BT with high $<Y>$ or a few
tens of millions of years for mass loading alone.

{It is important to notice in Fig. \ref{fig:vdrift} that the inward drift speeds increase as 
$\tau \rightarrow 0$. So, if we start with a ring of finite mass and radial extent, we expect
ML alone to empty the entire B ring onto the planet in a finite time, 
on the order of a factor of a few times the largest drift times given in Fig. \ref{fig:drift} in red 
(which are the times on which individual annuli of the rings drift all 
the way to Saturn). Global ring simulations performed in a companion paper verify this \citep{ED23}.
The case for BT alone is more complicated, because angular momentum is conserved, 
which means that some material must be left behind in orbit, but the bulk of mass elements
will flow into Saturn on the times suggested by the black curves for $<Y> = 10^5$. 
Also recognise that these drift times 
do not apply to narrow, dusty rings or sparse debris belts. The drift times presented here are 
indicative of how long a broad, dense ring can last. The true final behaviour of such a ring system
is a separate and interesting question. Perhaps it leads to a long-lived sparse system of 
dust belts and narrow ringlets.}

\subsection{Viscous Mass Inflow}
\label{subsec:visc}

It is interesting to compare the mass inflow in the rings
expected from viscosity with those derived above. 
Using the approximation of a steady-state 
viscous accretion disk with a constant kinematic viscosity $\nu$, 
the mass inflow rate is

\begin{equation}
\label{equ:Mdotvisc}
\dot{M}_{\rm{visc}} = 3\pi\nu\Sigma,
\end{equation} 

\noindent
when the disk is assumed to be Keplerian \citep[see, e.g.,][]{Har09}. 

{The proper value of viscosity in the rings is a complicated story, and expectations 
vary theoretically depending on the particle size distribution and whether or not gravitational
wakes occur \citep[see, e.g.,][]{Sch09,Sch16}. Values of the kinematic shear viscosity between 1 and 100 cm$^2$ s$^{-1}$
generally result, with low optical depth regions tending toward the lower end of the range
and higher optical depth regions toward the higher end (see discussions in E15 and E18).
We adopt values of 10 cm$^2$ s$^{-1}$ and 100 cm$^2$ s$^{-1}$, respectively, which 
probably overestimate the viscosity \citep[see, e.g.,][]{Tis07,Taj17}. For 
estimates of $\Sigma$ at low and high $\tau$, we use the same values adopted in 
Section \ref{subsec:age} below.}
 
{So, for} a low-$\tau$ region like the C ring, we get

\begin{equation}
\label{equ:Mdotvisclow}
\dot{M}_{\rm{visc}} = 0.2\; {\rm{kg}}\; {\rm{s}}^{-1}
\left(\frac{\nu}{10 \;{\rm{cm}}^2\; {\rm{s}}^{-1}}\right)  
\left(\frac{\Sigma}{2 \;{\rm{g}} \;{\rm{cm}}^{-2}}\right).
\end{equation} 

\noindent
For parameters more representative of the B ring,

\begin{equation}
\label{equ:Mdotvischigh}
\dot{M}_{\rm{visc}} = 50\; {\rm{kg}}\; {\rm{s}}^{-1}
\left(\frac{\nu}{100\; {\rm{cm}}^2\; {\rm{s}}^{-1}}\right)  
\left(\frac{\Sigma}{52\; {\rm{g}}\; {\rm{cm}}^2}\right).
\end{equation}

\noindent
These viscously driven inflow rates are orders of magnitude lower 
than we estimate for BT and mass loading in Eqs. (\ref{Mdotnumbertausmall}),
(\ref{Mdotnumbertaularge}), and (\ref{equ:MdotloadEKB}). 
Viscosity cannot explain the large mass inflow rates observed by Cassini at Saturn.

\section{Discussion}
\label{sec:Discussion}

There are two important implications of our 
results:  1) The high mass inflow rates through the rings for ballistic transport
and mass loading using plausible parameter choices are large 
enough to explain the large inflow rate into Saturn 
actually observed by Cassini at the innermost edge of the rings \citep{Wai18}. 
2) The age and future lifetime of the rings are
likely to be much shorter than the age of the Solar System. Both the age and lifetime are 
probably at most only a few hundred million years. The first three subsections 
below elaborate on these points. The last subsection offers some caveats 
and speculations.
 
\subsection{Mass Inflow to Saturn}
\label{subsec:inflow}

The mass inflow rates from the rings into Saturn's atmosphere as measured by various Cassini
instruments in the Grand Finale orbits cover a considerable range (Sec. \ref{subsec:finale}). 
Our general picture is that meteoroid effects drive a large inflow rate towards Saturn 
with the bulk of the inflow delivered equatorially through the rings. 
Various erosion processes along the way cause some significant flow into Saturn 
at high Saturn latitude in the form of ring rain as was measured by CDA, 
but these measurements are restricted to a narrow range of grain sizes. Such erosion mechanisms 
were suggested originally by \citet[][see review in D84]{CF70}, before there were many relevant observations.  
 
According to \citet{Hsu18}, the tens-of-nanometer grains {they detected with} CDA 
are ejected 
from the main rings at speeds of $\sim$ few $\cdot 10^2$ - $10^3$ m s$^{-1}$ with 50\% 
from the C ring, 40\% from the B ring, and the remaining fraction from the Cassini Division, 
A and D rings. The ejecta speeds for these small charged grains are larger than typical speeds 
assumed for the $\gtrsim$ micron-sized particles which dominate the impact ejecta size distribution 
in BT (CD90; D92; CE98; Sec. \ref{subsec:ejecta}), but are consistent with experimental 
\citep{AH96,SA12} and theoretical \citep{NC87} studies for smaller grains. 

It is informative to compare the total ejecta rate of $1800$ to $6800$ kg s$^{-1}$ 
derived by \citet{Hsu18} for these nanograins to the total ejecta production rate due to 
micrometeoroid impacts for all particle sizes across the rings. To estimate the
total impact ejecta emission rate from the main rings, 
we integrated the impact flux $\dot{\sigma}_{\rm{im}}$ over each ring 
assuming mean optical depths of $\tau = 0.1$ for the C ring and Cassini division, 
$\tau=2$ for the B ring and $\tau=0.5$ for the A ring to determine impact probabilities.
The result is 
 
\begin{equation}
\label{equ:totej}
\begin{split}
\dot{M}_{{\rm{ej}}} \sim 3.5\cdot 10^7\;{\rm{kg}}\;{\rm{s}}^{-1}\; \left(\frac{\dot{\sigma}_\infty}
{\dot{\sigma}_{\infty}({\rm{CDA}})}\right)\times \\
\left(\frac{<Y>}{10^5}\right)
\left(\frac{F_{\rm{g}}}{30}\right),
\end{split}
\end{equation}

\noindent
which is roughly the equivalent of a $\sim 20$ m radius object with the 
density of ice every second.
Thus the measured CDA {nanograin} flux, 
whose contribution spans the entire ring system, is probably no more than $\sim 10^{-4}$ of the total 
ejecta production rate. Of this only 18\% arrives at the planet as ring rain (the rest presumably returns 
to the main rings), and so the CDA nanograins represent only a small fraction of the typical impact yield $<Y>$ 
times the mass of the impactor. In fact, the total mass of such fast charged nanograins emitted in an impact 
are comparable in mass to the impactor itself. This minute component of small grains 
with very high ejection speeds appears to be consistent  with the correlation between 
fragment size and ejection speeds \citep{Sac15}.

As pointed out in Section \ref{subsec:finale}, the $100-370$ kg s$^{-1}$ of the CDA flux 
deposited at the ring rain latitudes in the form of charged, solid nanograin particles are 
apparently not sufficient to explain the $432-2800$ kg s$^{-1}$ required to account 
for the observed H$_3^+$ infrared emission pattern \citep{Odo19}. 
However, the silicate-to-ice mass fraction at the ring rain latitudes being as high as $\sim 30$\%
\citep{Hsu18} implies that a significant fraction of water has been liberated from these 
nanograins. If one imagines ``reconstituting'' the CDA grain population with sufficient
water so that it is similar to C ring material ($\sim 6$\% non-icy by mass), this gives an 
inflow of $\sim 470$ to $1750$ kg s$^{-1}$ of additional water, not too dissimilar from the 
estimate of \citet{Odo19}.\footnote{It should be acknowledged that this agreement may be 
coincidental because the CDA flux originates predominantly from the C and B rings, 
whereas the mass flux estimated from the \citet{Odo19} analysis comes overwhelmingly 
from the C ring.}

Most of the ejecta mass from impacts (Eq. [\ref{equ:totej}]) should be in larger grains, 
too large to be charged significantly, which is consistent with the flux measured 
from CDA (and MIMI) only representing a small fraction of the total inflow. The total 
midplane loss rate from these instruments is on the order of $225$ to $875$ kg s$^{-1}$, 
far less than the equatorial loss rate of $4800$ to $ 45000$ kg s$^{-1}$ measured by INMS,
which should sample the full range of sizes \citep{Wai18}. Because the origin of this material 
appears to be the D ring, it implies that the C ring must replenish the D ring material at a similar 
rate assuming a steady-state scenario. Likewise, the C ring must also be replenished 
from the B ring at a roughly similar rate.

From our analysis, for our nominal choice of parameters, the mass inflow rate through the 
B ring and the C ring due to ballistic transport and mass loading is consistent with the
observed mass inflow rates near Saturn as measured by INMS. 
For nominal parameters, the C ring mass inflow rate is
predicted to be a factor of several lower than that through the B ring. We find that even decreasing
one of the critical parameters, like reducing $<Y>$ to 10$^4$, 
which remains in a plausible range of parameters, the 
mass inflow rates for both rings stay within range of the lower bound
estimate for inflow from the INMS measurements.
Our main point here is that the observed 
mass inflow rates toward Saturn through the rings can, in principle, be sustained 
by ballistic transport plus mass loading without straining parameters. Some complications are addressed
below in Section \ref{subsec:spec}.

\subsection{The Age and Remaining Lifetime of Saturn's Rings}
\label{subsec:age}

We have argued for many years that the rings must be substantially 
younger than the Solar System owing to
consequences of meteoroid bombardment, based on both structural and
pollution effects (D92; D96; CE98; E15; E18). By determining 
some key parameters required to assess the importance of meteoroid 
bombardment, the Cassini Grand Finale has substantially strengthened the arguments for a 
young age (E18).
In addition, the measurement of the large inflow from the rings
into Saturn combined with a well-determined ring mass now also permits 
estimates for the future lifetime of the rings {(see also the discussion at
the end of Section \ref{subsec:load}).} 

First consider the age. The structural evolution of the rings due to BT occurs 
on time scales that vary from a few up to hundreds or thousands of $t_{\rm{G}}$ 
(Eqs. [\ref{equ:teegee}] and [\ref{equ:teegee2}]) 
depending on the structure being modelled (E18). The ages depend on the uncertain 
impact yield parameter $<Y>$ but are always less than the direct deposition time scale 
$t_{\rm{dd}} = \Sigma/\dot{\sigma}_{\rm{im}}$ for reasonable choices of $<Y>$.
The time $t_{\rm{dd}}$ itself does not depend on $<Y>$, and we expect  
the evolution of the rings  to their current state of pollution to be some fraction of $t_{\rm{dd}}$. 
Thus, age estimates based on pollution are more reliable measures of
absolute ring age than estimates based on BT structures alone.

As revisited by \citet{Kem23}, an 
absolute exposure time for the rings can be obtained by assuming they began as pure water ice and 
darkened to their current state \citep[][E15, E18]{CE98}. The rate per unit area at which the ring annulus 
at some $r$ with surface density $\Sigma$ is impacted is given by $\dot{\sigma}_{\rm{im}}$, so that 
the mass density of pollutant accumulated in a time $t_{\rm{pol}}$ is 
$\Sigma_{\rm{pol}} \simeq \eta \dot{\sigma}_{\rm{im}}t_{\rm{pol}}$, 
where the factor $\eta \le 1$ allows for the possibility that 
not all of the impactor survives the impact {as absorbing material} (CE98; E15). 
The mass fraction of pollutant after this time is $\Sigma_{\rm{pol}}/(\Sigma + \Sigma_{\rm{pol}}) 
\simeq \Sigma_{\rm{pol}}/\Sigma$. Measured values derived from Cassini RADAR for the amount of non-icy 
material in the rings are reported as volume fractions of the solid component $\upsilon$ 
\citep{Zha17a,Zha17b}. So, in terms of the mass fraction, 
$\upsilon \simeq (\rho/\rho_{\rm{pol}})(\Sigma_{\rm{pol}}/\Sigma) \ll 1$ 
where the mean solid density of a ring particle is $\rho = \rho_{\rm{pol}}\upsilon +
\rho_{\rm{ice}}(1-\upsilon)$. The density of ice $\rho_{\rm{ice}} = 0.9$ g cm$^{-3}$, and we take the 
solid density of the incoming pollutant to be 2.8 g cm$^{-3}$, which represents a volume mixture of 30\% 
carbonaceous and 70\% silicate material \citep{Woo17,Kem23}. The time to acquire 
the non-icy material then is

\begin{equation}
\label{equ:tpol}
t_{\rm{pol}} \simeq \frac{\rho_{\rm{pol}}}{\rho}\upsilon\frac{\Sigma}{\eta\dot{\sigma}_{\rm{im}}} 
\simeq \frac{\rho_{\rm{pol}}}{\rho}\upsilon\frac{t_{\rm{dd}}}{\eta}.
\end{equation}

For the B ring where $\tau$ is very large, {we found that} the {average} surface density 
{is 52} g cm$^{-2}$ based on the derived mass for the rings during the Cassini Grand Finale \citep[][Sec. \ref{subsec:Mdotnumbers}]{Ies19}, 
while the measured volume fraction of pollutant is $\sim 0.1$ to $0.5$\% \citep{Zha17b}, with the lowest 
values found in the very opaque B ring core. Using {the} median value for $\upsilon$ and {the average} $\Sigma$, 
we find that at $r_0 = 1.8$ R$_S$

\begin{equation}
\label{equ:tpolB}
\begin{split}
t^{\rm{B}}_{\rm{pol}} \approx 1.2\cdot 10^8\, {\rm{yrs}} \left(\frac{\Sigma}{52\,{\rm{g}}\,{\rm{cm}}^{-2}}\right)
\left(\frac{\dot{\sigma}_{\infty}({\rm{CDA}})}{\dot{\sigma}_\infty}\right)  \times \\
\left(\frac{30}{F_{\rm{g}}}\right)\left(\frac{r}{1.8 {\rm{R}}_S}\right)^{0.8}
\left(\frac{0.1}{\eta}\right)\left(\frac{\upsilon}{0.003}\right).
\end{split}
\end{equation}
 
For the C ring, we can do a similar calculation, except the low continuum optical depth, 
$\tau \sim 0.1$, means that the probability of absorption of a meteoroid, which is $\sim 0.2$, must be taken into account. 
A typical value for the surface density away from plateaus and the anomalous ``rubble belt'' in the 
middle C ring is $\Sigma \sim 2$ g cm$^{-2}$ \citep{Bai11,HN13,Zha17b}. 
Taking $\upsilon = 2$\% we get at $r_0 = 1.4$ R$_S$ a similar time scale

\begin{equation}
\label{equ:tpolC}
\begin{split}
t^{\rm{C}}_{\rm{pol}} \approx 1.2\cdot 10^8\, {\rm{yrs}} \left(\frac{\Sigma}{2\,{\rm{g}}\,{\rm{cm}}^{-2}}\right)
\left(\frac{\dot{\sigma}_\infty({\rm{CDA}})}{\dot{\sigma}_\infty}\right)  \times \\
\left(\frac{30}{F_{\rm{g}}}\right)\left(\frac{r}{1.4 {\rm{R}}_S}\right)^{0.8}
\left(\frac{0.1}{\eta}\right)\left(\frac{\upsilon}{0.02}\right).
\end{split}
\end{equation}

\noindent
In Eqs. (\ref{equ:tpolB}) and (\ref{equ:tpolC}), we have adopted a conservative value of $\eta = 0.1$ 
for grains with little or no volatile content \citep{Woo17,Kem23}. 

Both these time scales are {in agreement with} 
those derived in \citet{Kem23}. 
These time scales will be modestly longer if the 
rings began with a larger mass, 
as implied by the currently observed 
mass loss rates. Thus, given the complete suite of available Cassini data, the case appears fairly solid 
that the rings were formed 
{$\lesssim$ a few} 100 Myr ago, in agreement with arguments based on
pollution and structural models presented in previous works \citep[][CE98; E15; E18]{Doy89}. 

We establish this young ring age using the ring mass \citep{Ies19}, micrometeoroid flux at Saturn 
\citep{Kem23}, and the volume fractions of pollutants \citep{Zha17a,Zha17b}.
Combined with measurements of the mass loss rates into Saturn during the Cassini Grand
Finale, we can now also estimate the turnover time of the C and D rings
and the remaining lifetime of the rings as a whole. 

The mass loss rate {of nanograins} measured by CDA is a not a reasonable gauge for 
remaining ring lifetime because of the limited particle size range to which it is sensitive. On the
other hand, the mass flux of charged water products as calculated recently by \citet{Odo19}
may be more representative because the ring rain effect is probably caused by a contribution from all
particle sizes. Based on their estimates for the mid-latitude mass loss rate, these authors get 
a ring lifetime ranging from $\sim$ 150 to 1,100  Myr. with a likely value of 300 Myr,
just from the ring rain. 
This is an upper limit because it does not include the equatorial losses
detected in the INMS data \citep{Wai18}.

\citet{Wai18} do not estimate an overall remaining lifetime for the main rings, but they do
estimate the lifetime of the D and C rings from their mass inflow rates and obtain $7\cdot 10^3$ to 
$7\cdot 10^4$ yr for the D ring, and $\sim 7\cdot 10^5$ to $7\cdot 10^6$ yr for the C ring. Their
calculation for the D ring assumes that its mass is $\sim 1$\% 
of the $\sim 10^{18}$ kg C ring mass. 
Their D ring lifetime estimate is
based on the assumption that the D ring has a similar particle size distribution to the C ring so 
that the difference in surface density is simply the optical depth ratio between the D and C ring ($\sim 0.01$). 
However, meter-size particles dominating the mass in the D ring may not be a reasonable assumption. 
Particles sizes in the D ring of only tens of microns may be more appropriate \citep{Hed18b}. 
This would significantly lower the mass so that the current mass inflow rates would dissolve the D ring 
in only about a year. If so, this would tend to support the idea of episodic repopulation of the D ring 
from the C ring. 
\citet{Wai18} also posit that the D ring 
could be repopulated sporadically by large impact events like the one that tilted the D and C ring plane 
\citep{Hed11}.

From our own estimates of the mass inflow rates due to BT alone
in Eqs. (\ref{Mdotnumbertausmall}) and (\ref{Mdotnumbertaularge}), we can calculate the remaining
ring lifetime. Taking the current ring mass to be $\sim 0.4$ Mimas masses \citep{Ies19}, 
or $\sim 1.5\cdot 10^{19}$ kg, we find that the range of times for the remaining mass of the rings to 
fall into the planet is {$\sim$ 15 to 40 Myrs} for the case when $<Y>\; =10^5$, or {$\sim$ 150 to 400 Myrs} for
$<Y>\; =10^4$. This assumes that the mass inflow rate remains constant over that time.
The lifetime estimate would be shorter at the tens of percent level or less,
depending on parameter choices,  
if we include contributions due to mass loading and meteoroid torques. 
They can also be shorter if the slope of the ejecta
velocity distribution is flatter so that more ejecta mass is placed at higher speeds
or if there is an accentuation of the inward drift rate of material 
due to the \citet{Goe86} mechanism (see Section \ref{subsubsec:EMtorques}). 
Our lower bound estimate for the lifetime is naturally similar to what would be inferred from
the upper bound for the mass inflow measured by INMS 
because BT produces mass inflow rates of the 
same order as observed for our nominal parameter choices. {The future lifetimes estimated
here based on ring mass and inflow rates are similar 
within a factor of a few to the inward drift times in 
Fig. \ref{fig:drift} of Section \ref{subsec:load}, as they should be.} 

The lower bound estimate of {15 Myr} seems exceedingly short, perhaps suggesting
that the yield $<Y>$ or other parameters vary across the rings and/or $<Y>$ is
generally lower than what we adopt. Even if
the influence of BT alone is not as great as computed here, 
our results show that mass loading alone, which is
{not far off} from the lower bound measured by INMS, 
represents a relatively firm upper bound on the ring lifetime, similar to
the upper bound estimated above for the low-$<Y>$ BT case, namely {$\sim$ 400 Myr}.
It could also be that, if the upper estimate of the mass inflow rate by INMS 
is correct, the inflow just happened to be exceedingly high at the time of observation 
due to some recent event \citep{Wai18}.

\subsection{Initial Ring Mass}
\label{subsec:InitialMass}
 
The mass inflow rates rates from our BT analysis indicate that the ring mass was 
substantially larger in the past. Given our estimate for the current ring age of $\sim 10^8$ years 
based on pollution, this would imply that the initial ring mass was greater by $\sim$  few tenths
to a few Mimas masses. The inclusion of mass loading increases these numbers by tens of percent. 
{So, as we argue in more detail in our companion paper \citep{ED23}}, 
the deduced initial ring mass is consistent with a recent origin scenario 
in which the rings start significantly more massive 
($\sim$ one to several Mimas masses) and are polluted by micrometeoroids as they evolve to their 
current mass and darkened state. 
 
The mass inflow rates measured by Cassini, as well as those that we calculate here, 
suggest that relatively massive rings are ephemeral when subjected to 
micrometeoroid bombardment\footnote{BT may not be so influential in less massive and sparse ring 
systems such as Uranus and Neptune in which the rings and arcs are radially thin.}. 
Previously, for evolution by viscosity alone, it was found that a massive ring evolves until 
its surface density drops to the point where the viscous spreading time 
becomes exceedingly long \citep{Cha09}. The ring mass then approaches an asymptotic limit 
\citep{Sal10}. The current ring mass is close to that asymptotic limit. However, this result only 
applies in the absence of other mass loss mechanisms. Given the results of observations and our 
analysis here, it seems that the ring mass being similar to the \citet{Sal10} theoretically-derived asymptotic 
mass may be a coincidence. 

\subsection{Caveats and Speculations}
\label{subsec:spec}

\subsubsection{The C and D rings}
\label{subsubsec:CDrings}

Eqs. (\ref{Mdotnumbertausmall}), (\ref{Mdotnumbertaularge}), and 
(\ref{equ:MdotloadEKB}) plus Figures \ref{fig:mdotin} to \ref{fig:mdotload} show that, at 
high $\tau$ for nominal parameters, mass loading and ballistic transport can cause mass inflow
rates through the B and C ring consistent with observations. Even though 
the mass inflows for both mechanisms are somewhat weaker  
at low $\tau$ ({\it i.e.}, at decreasing surface density), the radial inward drift speeds due to
these mechanisms asymptote to large nonzero values (Fig. \ref{fig:vdrift}) leading to short
loss time scales for ring particles in the C ring and inner B ring (Fig. \ref{fig:drift}). 
The low-$\tau$ C ring does not extend all the way down to Saturn, but, for our nominal
set of parameters, the mass inflow rate is roughly sufficient to account for the amount of mass
that is pouring into Saturn's atmosphere. On the other hand, if $<Y>$ or $\dot{\sigma}_\infty$ 
are considerably smaller than we assume, the C ring mass inflow rate might not be 
sufficient, especially for the case where the upper estimate for the observed mass inflow rate
is correct. Then something important may be missing. 

$\dot{M}_{\rm{bt}}$ in the C ring can be made to match the full range of the observed 
mass inflow rates by adopting yields between $\sim 10^4$ and $10^5$. As we discussed in Sec. 
\ref{subsec:parameters}, structural models of the inner B ring edge and its transition into the 
low-$\tau$ C ring require yields $\gtrsim 10^6$ for a cometary meteoroid source in order to 
keep the edge relatively sharp, maintain roughly its observed width, and produce a ramp 
(D92; E15). The updated BT models using the EKB meteoroid source determined by CDA \citep{Kem23} 
demonstrate the same morphology but only require yields of $\gtrsim 10^5$ to maintain the B ring 
edge and ramp \citep[][E18]{Est16}. So there is indirect evidence that the BT mass inflow may be 
on the order suggested by Eq. (\ref{Mdotnumbertausmall}). However, it may be the case that $<Y>$
is smaller away from the inner edge and/or other properties of the ring particles conspire
to reduce the mass inflow rate.

Alternatively, inflow from the D ring into Saturn could be episodic
or highly variable with a duty cycle of a third to a tenth, and Cassini happened to sample a much 
higher than average inflow. How
the C ring's structure changes with time may also have an effect on the mass inflow
rate. Structures such as plateaus may be long-lived even if material is drifting through them,
or they may change and/or drift inwards on relatively short time scales. There are some gaps
in the C ring which might serve temporarily as barriers to mass inflow, but the throw distances
of ejecta vary from 10's up to 1000's of km, so that some ejecta can ``hop'' over gaps in both
directions. They can hop inwards as well as outwards 
because some fraction of impact ejecta are always thrown inwards (CD90). 
Curiously, most of the gaps in the C ring have optical depth spikes at their outer edges, possibly a pile
up of material that would be expected if there was a net inward drift rate of material.
Properly evaluating how the mass flow behaves under these structural conditions will require 
future detailed numerical simulations.

The mass loading result is independent of any assumptions about the nature of the transport, but 
the stronger BT mass inflow rates are computed using the simple model of a constant, nearly steady-state 
ring. As noted in the previous paragraph, the real rings exhibit gaps, sharp edges, and structure on a 
variety of scales, so the quasi-steady solution applies at best only as an overall ballpark estimate. 
While many ejecta can leap across much of the small scale structure, the assumption of a uniform ring 
clearly breaks down. While the A and B rings probably experience a large overall mass inflow
rate, on the order of our estimates, especially across their more uniform regions, there is clearly a 
need for global BT simulations that include more complexity. 

Our inflow model is incomplete in the sense that our formulation of ML and BT does
not apply to the D ring itself, which is extremely optically thin and possibly dusty, nor does it
address what happens in the gap between the D ring and Saturn, which is wider than a
typical ejecta throw distance. A complete
discussion goes beyond the scope of this paper, but we point out some considerations. 
As material flows into the inner C ring, the higher velocities of meteoroids and ring particles 
make the impacts more catastrophic, with more 
smaller particles, vapor, and ions being produced. Ejecta velocities are likely to be higher. 
Once a ring begins to suffer net erosion rather than loading, typical particle sizes are likely to shrink. 
Small charged ejecta particles and ions will move along magnetic field lines into Saturn's atmosphere. 
Larger charged ejecta lose angular momentum to Saturn's magnetic field 
(see Section \ref {subsubsec:EMtorques}). The occurrence of ring rain, which falls at moderate 
latitudes, presumably along magnetic field lines, demonstrates that processes like these are indeed 
occurring in the D and inner C ring \citep{Hsu18,Wai18,Mit18}. Ring particles eroded and shattered to smaller sizes will 
become subject to plasma drag as they approach the upper layers of Saturn's atmosphere
\citep{Hed18b}. Detailed theoretical or computation studies of the fate of ring material as it 
gets close to Saturn is difficult but certainly worthy of further research.

\subsubsection{Composition of the Inflow}
\label{subsubsec:compinflow}

The composition of the inflowing material as seen by CDA appears to be consistent with both C ring
composition and the amount of charged water products needed to account for the ring rain phenomenon in
Saturn's atmosphere, if one reconstitutes for the missing water (Sec. \ref{subsec:inflow}).
However, the measured INMS composition has a non-icy mass fraction of $\sim 75$\% far in excess of
bulk ring composition; moreover, the INMS flux is restricted to a narrow equatorial
band, and thus must apparently originate from the midplane with inner C or D ring material providing
the source \citep{Wai18}. Indeed, if one were to reconstitute the inflow to mean C ring composition
($\sim 2$\% volume fraction), it would imply a mass flux of $\sim 6\cdot 10^4 - 5.6\cdot 10^5$ kg
s$^{-1}$. BT in principle could provide such large mass influxes for impact yields $\gtrsim 10^6$ 
(Sec. \ref{subsec:Mdotnumbers}), but then where does the water go?
It may be possible that a higher fraction of ejecta could be vaporized due to much higher micrometeoroid
impact velocities in the D and inner C ring deep in the planet's potential well, which subsequently may 
recondense on the rings, but it would seem unlikely that such a large fraction of vapor ($> 90$\%) would be
produced in this way.
If the latter were the case, it might 
suggest that meteoroid bombardment is actually purifying the ring composition, and not polluting it.
For now, we consider this idea entirely speculative
and hope to address it more extensively in a future paper. 
The fact that there is an order of magnitude contrast in the level of
darkening between the C ring and B ring, which is consistent with 
direct deposition of micrometeoroid pollutants (E15, E18), 
argues against
a purification process. It is made even more implausible given that BT tends to produce
a more uniform distribution of non-icy material if allowed to act over long 
enough time scales.  

Nevertheless, we note that the composition of the volatiles as observed by INMS 
appears similar to that of comets. Perhaps material from the
icy progenitor moon (or moons) that formed the rings was composed of mostly unprocessed, primitive
ices \citep{Wai18}. Cassini INMS found, from flying through plumes, that 
Enceladus and comets are chemically very similar \citep{Wai17}, which suggests that the material
that formed Enceladus (and likely all the Saturnian moons) came from heliocentric bodies. This
is consistent with the idea that the primary source of solids for the subnebulae of giant
planets like Jupiter and Saturn is ablation and capture of planetesimals crossing their
circumplanetary disks \citep[][also see \citet{MG04}]{EM06,Est09,Mos10}.

Another possibility, perhaps the simplest, 
is that, given the short lifetime of the C ring (see Figure 6), the admixture
of Centaur material into the middle C ring in the last few {tens of} million years
\citep[as suggested by][]{Zha17b}\footnote{{These authors also discuss the possibility that the rubble could be the last remnants of a ring progenitor moon's core.}}
may now be dominating what is accreting into Saturn. 

\subsubsection{Satellite Torques}
\label{subsubsec:satellites}

It has been known for some time that torques exerted through orbital resonances 
with Mimas and the various inner and embedded satellites have a substantial effect 
on ring structure, producing density and bending waves and sharp edges
\citep[e.g.,][]{Lis81,Bor89,HS92}. 
These torques also transfer angular momentum from the rings to external
satellites and produce short orbit evolution time scales for some satellites \citep[e.g.,][]{PS01}.  
The radial drift due to pure mass loading does not involve angular momentum
transport, but the inward drift of mass in the steady-state BT solution is caused 
by the outward transport of angular momentum through the predominance of prograde 
ejecta. A natural question to ask is where that angular momentum goes in a real ring
with edges. 

The so-called ``box car" evolution, an initially constant-$\tau$ ring between two sharp 
edges, is illustrative. Figure 10 of D89 shows that, for prograde ejecta, even though the 
entire box car drifts inward,  a peak develops near the outer edge to conserve angular
momentum. In real rings, which do not extend outward infinitely at constant $\tau$, there
will be similar effects due to edges or boundary conditions. For a free outer boundary,
the BT-driven mass inflow will restructure the rings
to conserve angular momentum. How this will affect the rings globally over long times will 
be the subject of future research. 

What we can say is that the various satellite
torques, as given in Table 2 of \citet{Taj17}, are insufficient by several
orders of magnitude to remove the outward angular momentum flux due to BT. The
satellite torques are typically $\sim$ few$\cdot 10^{13}$ Nm, while a BT mass inflow  
{of only $\sim 3\cdot10^{4}$ kg s$^{-1}$} implies angular momentum outflow 
resulting in an equivalent torque 
{approaching $\sim 10^{16}-10^{17}$ Nm}.
The satellite torques simply cannot keep up with the
BT-driven angular momentum outflow, although they can balance angular momentum 
outflow due to viscosity, as in \citet{Taj17}, 
because the viscous flow is orders of magnitude smaller. The 
smaller angular momentum outflow for viscosity is reflected in the correspondingly lower
viscous mass inflows, Eqs. (\ref{equ:Mdotvisclow}) and (\ref{equ:Mdotvischigh}), calculated for the
steady-state case (see Section \ref{subsec:visc}).
We currently draw no simple conclusion about this difference between the BT-driven 
angular momentum outflow and satellite torques. As mentioned in Section \ref{subsec:future},
once we have included proper ballistic transport in global simulations,  we
will study what happens near outer edges.

While not now sufficient to balance the net BT angular momentum
transfer, satellite torques still probably dominate locally for balancing 
viscous edge spreading on very short length scales. 
For the expected ejection velocities of between a few to a few times 100 m s$^{-1}$, 
the radial ejecta throw distances range from a few 10's to a few 1,000's km. So BT transfer 
is probably not important over extremely sharp edges nor over the relatively short wavelengths 
of the spiral structure induced by resonances. The huge disparity of the dynamic and 
gross erosion times also implies that BT should not affect the physics of spiral density 
waves. On the other hand, 
it will be interesting to see, in future BT simulations, how BT
affects regions of 100's to 1000's km width inside sharp edges.

In some global ring evolution scenarios, 
a proto-Mimas may once have been much closer to the outer A ring edge. There may have
been an early, more massive phase of the rings when considerable angular momentum 
was transferred to Mimas 
leading to the formation of the Cassini
division \citep{Bai19}. Looking to the 
future, the current rings may be stuck with close to their current total angular momentum, 
even as they lose mass. The end state of such an evolution needs to be addressed.

\subsubsection{Electromagnetic Effects}
\label{subsubsec:EMtorques}

It seems likely that ring rain and other high latitude mass inflows from the rings into Saturn 
are due in part to grain charging and electromagnetic effects (see Section \ref{subsec:finale}), 
which cause tiny charged grains or ions to move along Saturn's magnetic field lines. 
As indicated by Eq. (\ref{equ:totej}), the production rate of ejecta by meteoroid bombardment 
is prodigious. It would take only a small fraction of a percent of ejecta to be small enough and 
to become charged enough in order to explain ring rain and the other high latitude inflows. 

\citet{Goe86} pointed out that somewhat larger charged grains would not reach 
the planet and would instead lose (gain) 
angular momentum as they travel due to magnetic torques exerted inside (outside) 
the corotation radius ($\sim$ 1.86 ${\rm R}_S$) with Saturn's magnetic field.
When these grains lose their charge or otherwise rejoin the ring material,  they would
be reabsorbed in the rings at locations with different specific angular momentum. 
The interesting point for application to mass inflow is that this could enhance 
the inward mass inflow by providing another angular momentum loss mechanism 
over the C ring and the inner two-thirds of the B ring. \citet{Goe86} estimated that the 
torques were comparable to or stronger than those due to viscosity. A full re-evaluation 
involves using modern Cassini data to estimate production rates of charged grains 
of the appropriate size and the radial distances they are likely to travel. This goes beyond 
the scope of the current paper but might be a fruitful avenue of future research. 

As a follow up to \citet{Goe86}, \citet{GM88} showed that the Goetz et al. 
effect could produce an instability leading to wavelike structure, similar to the BT linear 
instability \citep[D95;][]{Lat12,Lat14a,Lat14b}. As a possible indication that the Goertz et al. mechanism
really operates in the rings, we point out that the Goertz and Morfill instability offers a 
plausible explanation for why the undulations in the C ring have a long wavelength and are 
sustained at a nonlinear amplitude without propagating quickly into Saturn, as would be expected 
if BT alone were the cause \citep[][E15; E18]{Lat12,Lat14a,Lat14b}. 

\subsubsection{Non-Detection of Systematic Inward Drift} 
\label{driftsnotseen}

{Over the roughly forty years between the Voyager missions and the Cassini Grand Finale, 
the inward drift speeds for the C ring in Fig. \ref{fig:vdrift} should lead to radial material shifts of 
between about 0.04 km for ML alone and about 1.2 km for BT with $<Y> = 10^5$. In principle, 
the BT drifts are large enough to detect if appropriate ring features can be identified to track the flow. 
\citet{Fre17} performed a comprehensive study of the radial positions of over a hundred relatively
sharp ring features distributed across the rings as determined by stellar occultations. 
The derived radii of most features do not 
differ by more than a few hundredths of a km between fits that include only Cassini data and 
fits that include Voyager and Earth-based data. This suggests little overall 
drift detected in the chosen features. When compared with the pre-Cassini radial scale 
determined by  \citet{Fre93}, 
the subset of individual features common to both studies 
do show radius discrepancies of up to several kilometres, outside the formal error bars, 
but these are as often positive as 
negative and not indicative of overall inward drift.}

{It is incorrect to conclude from this that $<Y>$ must be considerably less than 10$^5$. The
problem is that ML and BT do not necessarily cause identifiable ring features to co-move with the
inward flow.}  

{To illustrate this, consider the two long BT simulations for the inner B ring edge region 
in Figs. 8 of E15 using $<Y> = 1.5\cdot10^6$ 
and cometary impactors ($F_{\rm{g}} \approx 3$). For one simulation (dark line), 
the inward motion of the inner edge feature is less than about 10 km over 
7$\cdot$10$^6$ yr.  For the other simulation (red line), which differs only in assumptions about
ring opacity, the edge moves inward by $\sim$ 100 km in that time. These inward shifts are 
orders of magnitude slower than the inward drift of mass elements in the simulations. The 
relatively sharp inner B ring edge is maintained by BT 
but is more like a waveform than a material structure. 
Mass flows inward through the edge much faster than the edge itself moves. 
The set of shorter simulations with  $<Y> = 3\cdot10^5$ in Fig. 9 of E15
re-enforces the point that mass inflow does not have to result in inward motion of 
features.  All the simulations start with an inner B ring edge centered 
at 1.52 $R_{\rm s}$, but the edge can move either inward 
or outward, depending only on the exponent of $n$ in the ejecta speed distribution 
(see Section \ref{subsec:ejecta}).}

{It is interesting to note that some of the largest differences 
between the \citet{Fre93} and \citet{Fre17} fits, as shown in the 2017 paper's Table 5,  
are associated with the edges of the plateaus in the 
C ring. As we discuss in E15 and E18, ballistic transport may play some role in the evolution
of these features, even if it is not their cause. Intriguingly, 
in the work by the French group, all the inner edges of plateaus shift 
systematically outward by about 1 to 2 km and the outer edges shift inward by about 1 to 2 km. 
The magnitude is on the order of the BT radial drift in our Fig. \ref{fig:vdrift}, 
but the opposite signs are perplexing. These shifts do not directly reflect a 
systematic inward flow but could reflect the magnitude of ballistic transport effects
acting on the plateaus.}

{Overall, the work by French's group on radial scales in the rings
is inconclusive with regard to the presence of the large inward drifts suggested by 
Fig. \ref{fig:vdrift}. BT simulations of edge or gap features which are not maintained by
BT itself might be enlightening (see Section \ref{subsec:future}).} 

\section{Conclusions}
\label{sec:conclusion}

\subsection{Mass Inflow Rates}
\label{subsec:inflowconcl}

The main conclusion of this paper is that, in light of new determinations of key
parameters by Cassini, ballistic transport and mass loading due
to meteoroid bombardment can plausibly explain the magnitude of the mass inflow 
from the rings into Saturn as observed during the Grand Finale. 
For our nominal choices of parameters, we show that these mechanisms 
can drive mass inflow through the B and C rings with rates that fall within the range determined 
by INMS \citep{Wai18}. 

The mass loading inflow rates computed in Section \ref{subsec:load} 
depend only on the product of the Saturn Hill radius meteoroid influx $\dot{\sigma}_\infty$ 
and the gravitational focusing factor $F_{\rm{g}}$. This product is now determined 
by CDA measurements of dust impacts
\citep{Kem23}. So the mass loading results are pretty secure. At moderate to 
high values of $\tau$, mass loading alone is {comparable to (a factor of two smaller)}
the lower range of the INMS inflow measurement.

For ballistic transport, we use the approximate steady-state solution 
developed in Section \ref{sec:inflow} to estimate mass inflow rates. In addition
to the factors $\dot{\sigma}_\infty$ and  $F_{\rm{g}}$, the ballistic transport mass inflow 
(Section \ref{subsec:Mdotnumbers}) depends on the less well-determined factors 
of typical impact yield $<Y>$ and average normalized ejecta speed $x_b$. 
The ballistic transport inflow rates are roughly compatible with the lower and upper
estimates of the observed mass inflow rates reported by INMS 
for impact yields $<Y>\; \approx 10^4$ to $10^5$, respectively. While 
it is unclear how well the steady-state solution applies to the real rings locally,
it probably gives a useful global estimate for inflow. The meteoroid-driven
mass inflows are several orders of magnitude greater than mass flows due
to viscosity (Section \ref{subsec:visc}).

Regardless of the various uncertainties, discussed in some detail in 
Section \ref{subsec:spec}, this paper shows that
mass should be driven inward toward Saturn in the B ring at rates high enough
to explain the observed mass inflows into Saturn. Even though mass inflow rates 
in the C ring can also be large enough under reasonable assumptions, 
the situation in the C ring is not as clear because it contains a lot of unexplained complex 
structure and the detailed loss mechanisms operating close to Saturn are not yet 
well enough understood.

\subsection{Ring Age, Lifetime, and Initial Mass}
\label{subsec:agelifemass}

The mass inflow rates driven by ballistic transport and mass loading do not in
themselves dictate a ring age. Instead, in Section \ref{subsec:age}, we used modern data
to estimate a ring age due to pollution by meteoroids of only $\sim$ {170 Myr}. Also
in Section \ref{subsec:age}, we showed that the ballistic transport mass inflow rates in 
Eqs. (\ref{Mdotnumbertausmall}) and (\ref{Mdotnumbertaularge}) 
suggest future ring lifetimes of as little as {15} to as much as {400 Myr} using
the specific examples of impact yields of $10^4$ and $10^5$. These lifetimes are reduced 
by tens of percent when mass loading is included. The range of 
measured mass inflow rates from INMS, which are consistent with our results here, if
steady, are large enough to deplete the C ring in only $\sim 0.7$ to 7 Myr.

As discussed in Section \ref{subsec:InitialMass}, the ballistic transport mass inflow rates for 
nominal parameters operating over our estimated $\sim$ {120 Myr} age for the
rings suggest that $\sim$ a few tenths to several Mimas masses have already been lost
into Saturn. Adding mass loading increases the numbers modestly. 

\subsection{Future Work}
\label{subsec:future}

Although it will probably make differences of a factor of a few
at most, the CD90 analysis of focusing and aberration effects for meteoroid 
impactors with a cometary source needs to be repeated for the EKB source
determined by \citet{Kem23}. E18 estimated the changes with 
simple approximations but did not redo the CD90 analysis in detail. 
The various $\tau$-dependences for impactor absorption 
probabilities, ejecta emission rates, and mass loading drift are likely to be 
affected by the change from a cometary to an EKB source. For instance, 
the $\tau_0$ in Eq. (\ref{equ:Rtau}) will probably be smaller
due to the greater aberration of the EKB impactors leading to increased slant
path optical depths experienced by the meteoroids at the rings. It is also likely that 
meteoroid torques  will play a larger role for EKB impactors for similar reasons.
So, we plan to redo the CD90 analysis for EKB projectiles and derive 
proper ejecta distribution functions and $\tau$-dependencies.

It is clear, by comparing the viscous mass inflow rates 
in Eqs. (\ref{equ:Mdotvisclow}) and (\ref{equ:Mdotvischigh})
with the ballistic transport rates in 
Eqs. (\ref{Mdotnumbertausmall}) and (\ref{Mdotnumbertaularge}) and the mass loading
rate in Eq. (\ref{equ:MdotloadEKB}) that the inclusion of mass 
loading and ballistic transport in 
global ring simulations is necessary. 
{In our companion paper \citep{ED23}, we show}
that adding just mass loading to global viscous evolutions 
leads to finite lifetimes for massive rings rather than the asymptotic
mass found in purely viscous simulations \citep{Sal10}.

Adding ballistic transport correctly to global simulations is computationally 
involved but should shorten the expected lifetimes further and have interesting 
implications for the current mass distribution of the rings because of the outward 
angular momentum transport involved. Global pollution evolutions could be run
simultaneously, and the possibility of a changing composition due to selective
retention of water during the inflow process should be considered.
Once global simulations with proper ballistic transport become possible, it will be 
interesting to include  satellite torques, especially in scenarios where a 
proto-Mimas forms outside the Roche radius and then recedes due to 
resonant torques. Other mass and angular momentum transport processes, 
like the \citet{Goe86} mechanism could also be considered. In addition to
global simulations, localized ballistic transport simulations will be 
needed to study how mass inflow is affected by gaps and 
other ring structures, especially in the C ring. 
 
\vspace{0.0in}
\begin{center}
{\bf Acknowledgements}
\end{center}
\vspace{0.0in}
We thank Josh Colwell, Jeff N. Cuzzi, Joe Burns, and Glen Stewart for useful discussions. 
This work was supported by a grant from NASA's Cassini Data Analysis Program (PRE) and 
a Cassini IDS grant (PRE) to Jeff N. Cuzzi. We are also extremely grateful to two
anonymous referees for thorough and thoughtful reviews that led to extensive revisions of the
manuscript.




\vspace{0.2in}
\bibliographystyle{model2-names.bst}\biboptions{authoryear}
\bibliography{my2.bib}







\end{document}